\documentclass[%
superscriptaddress,
nofootinbib,
 amsmath,amssymb,
 aps,
pra,
floatfix,
]{revtex4-2}
\usepackage{array,multirow,graphicx,makecell}
\usepackage[font=small,labelfont=bf,justification=justified]{caption}
\usepackage{subcaption}
\usepackage{ftnxtra}
\usepackage{adjustbox}
\usepackage{booktabs}
\usepackage{dcolumn}
\usepackage{hhline}
\usepackage{fontawesome}
\usepackage{bm}
\usepackage{float}
\usepackage[colorlinks,citecolor=blue,urlcolor=blue,linkcolor=blue]{hyperref}
\usepackage{color}
\usepackage{orcidlink}  
\usepackage{graphicx}
\usepackage{physics}
\usepackage{tabularray}
\usepackage{threeparttable}
\usepackage{comment}
\usepackage{float}
\usepackage{bm}
\usepackage{natbib}

\graphicspath{{./}{fig/}}

\begin{document}


\title{Two Cosmological Histories in a New Exponential Infrared $f(T)$ Teleparallel Gravity: Implications for the Hubble Tension and the Low-$\ell$ Planck TT--EE Tension}

\author{Mohamed Bedair\,\orcidlink{0000-0001-6446-1613}}\email{mohamedbedair@du.edu.eg}
\affiliation{Department of Mathematics, Faculty of Science, Damietta University, Egypt}
\begin{abstract}

We investigate a particular case of the New Exponential Infrared $f(T)$ teleparallel gravity model with the same number of free parameters as the standard $\Lambda$CDM model. Remarkably, due to the two real branches of the Lambert function $W$, the model admits two distinct cosmological solutions, denoted by $f^{(I)}$CDM and $f^{(II)}$CDM. The former exhibits a late-time phantom crossing, whereas the latter predicts an effective dark-energy density evolving from negative to positive values.

Using a Markov Chain Monte Carlo analysis of the Planck CMB, DESI BAO, and Type Ia supernova data, we find that the $f^{(II)}$CDM branch develops a bimodal posterior distribution in the primordial parameters $A_s$ and $\tau_{ reio}$ for the Planck-only dataset. We show that this bimodality originates from the tension between the low-$\ell$ TT and EE likelihoods, which becomes pronounced in models exhibiting a significantly enhanced late ISW contribution relative to the $\Lambda$CDM model. Moreover, the inclusion of background observations drives this branch toward cosmologically unacceptable parameter values, rendering it observationally unviable.

The $f^{(I)}$CDM branch, on the other hand, remains observationally viable, yielding $H_0=73.43\pm0.69~\mathrm{km\,s^{-1}\,Mpc^{-1}}$ for the Planck-only dataset and $H_0=72.55\pm0.32~\mathrm{km\,s^{-1}\,Mpc^{-1}}$ after including DESI, substantially alleviating the Hubble tension. Although this branch provides a slightly better fit than $\Lambda$CDM for  Planck-only data, the addition of background observations reverses this preference in favor of the standard $\Lambda$CDM model. These results show that the cosmological viability of this model critically depends on the choice of the Lambert $W$ branch, with only the $f^{(I)}$CDM solution remaining compatible with current cosmological observations.

\end{abstract}
\maketitle
\newpage
\section{INTRODUCTION}
\label{sec:intro}

The standard cosmological model, $\Lambda\text{CDM}$, built upon General Relativity (GR) with a cold dark matter component and a cosmological constant $\Lambda$, has achieved remarkable success in describing a wide range of observational data, from the Cosmic Microwave Background (CMB) anisotropies to the Large-Scale Structure (LSS) of the Universe \cite{Planck:2018vyg, Menci:2024hop}. Despite its observational robustness, $\Lambda\text{CDM}$ faces severe theoretical and empirical challenges \cite{Nojiri:2017ncd}. Theoretically, it suffers from the cosmological constant problem and the fine-tuning issue associated with the vacuum energy density. Observationally, persistent tensions have emerged between early- and late-universe measurements \cite{Cai:2026swf}. Most prominently, the $H_0$ tension a statistically significant discrepancy exceeding $5\sigma$ between the Planck CMB-inferred value of the Hubble constant ($H_0 \approx 67.4 \text{ km s}^{-1}\text{Mpc}^{-1}$) and local distance ladder measurements, such as those by the SH0ES collaboration ($H_0 \approx 73.0 \text{ km s}^{-1}\text{Mpc}^{-1}$)---suggests potential new physics beyond the standard model \cite{Riess:2016jrr, Menci:2024hop}.

To address these tensions without invoking ad hoc dark energy components, modified theories of gravity have garnered considerable attention \cite{Cai:2026swf, CANTATA:2021asi}. Among these, $f(T)$ teleparallel gravity offers a compelling alternative framework \cite{Bahamonde:2021gfp}. By replacing the Levi-Civita connection of Riemannian geometry with the torsion-based Weitzenböck connection, teleparallel gravity attributes gravitational interactions to the torsion scalar $T$ rather than curvature $R$ \cite{Heisenberg:2023lru, Aldrovandi:2013wha}. Generalizing the teleparallel action to an arbitrary function $f(T)$ yields second-order field equations, thereby avoiding Ostrogradsky instabilities that frequently plague higher-derivative curvature-based theories \cite{Li:2010cg, Bahamonde:2021gfp}.

Recently, a novel class of infrared exponential $f(T)$ gravity models was introduced to explore late-time dark energy dynamics without introducing additional free parameters beyond the baseline $\Lambda\text{CDM}$ set \cite{Awad:2017yod, Hashim:2020sez,Hashim:2021pkq}. A key feature of these models stems from the double-valued nature of the Lambert $W$ function, which governs the relation between the present torsion density parameter and the model parameter $\beta$ \cite{Akarsu:2024nas}. Consequently, the theory admits two distinct cosmological histories corresponding to the two real branches, $W_0$ and $W_{-1}$ \cite{Akarsu:2024nas, Hashim:2026yoy}.

In this work, we present a comprehensive observational and theoretical investigation of a non-degenerate case within this infrared exponential $f(T)$ framework, defined by $f(T) = (T + \sqrt{T_0 T})\exp(\beta T_0 / T)$ \cite{Hashim:2026yoy, Bedair:2026M4}. We examine the resulting two branches: the primary $f^{(I)}\text{CDM}$ branch, which exhibits late-time phantom-crossing behavior, and the secondary $f^{(II)}\text{CDM}$ branch, which features a sign-changing effective dark energy density evolving from negative to positive values \cite{Akarsu:2024nas, Gomez-Valent:2025mfl}. Utilizing a full Markov Chain Monte Carlo (MCMC) analysis incorporating Planck 2018 CMB, DESI DR2 BAO, CMB lensing, and multiple Type Ia Supernovae compilations, we evaluate the observational viability of both branches, analyze their impact on the $H_0$ tension, and uncover a novel bimodal posterior distribution associated with the low-$\ell$ TT-EE likelihood tension in $f^{(II)}\text{CDM}$.

The paper is organized as follows: In Sec. \ref{fid}, we review the foundational framework of $f(T)$ gravity and its cosmological field equations at the background and linear perturbation levels. In Sec. \ref{sec:model}, we introduce the exponential infrared model and derive its two cosmological branches via the Lambert $W$ function. Sec. \ref{sec:phase} provides a phase portrait analysis of the fixed points and asymptotic limits. Sec. \ref{sec:data} describes the observational datasets and MCMC setup. In Sec. \ref{sec:res}, we present the observational constraints, model comparison, and detailed discussion of parameter degeneracies. Finally, Sec. \ref{sec:con} summarizes our main conclusions.
 \section{\texorpdfstring{$f(T)$}{} Theory and  Cosmology }\label{fid}
In this section, we briefly review the mathematical framework of teleparallel geometry, in which the torsion tensor, rather than the curvature tensor, characterizes the underlying spacetime geometry. We then present the field equations of $f(T)$ gravity, obtained by replacing the torsion scalar in the Einstein--Hilbert action with an arbitrary function $f(T)$. Finally, we summarize their application to cosmology at both the background and linear perturbation levels.

\subsection{ \texorpdfstring{$f(T)$}{} Theory }
In any spacetime manifold, there exist two independent structures necessary to define the geometric properties of the space. The first is the metric tensor $g_{\mu \nu}$, which defines the norm and angles between tangent vectors in the tangent space. The second is the affine connection $\Gamma^{\alpha}_{\phantom{\alpha}\mu \nu}$, which defines the parallel transport of vectors. Three fundamental geometric tensors can be defined in terms of these structures: the non-metricity tensor $Q_{\alpha \mu \nu}$, the torsion tensor $T^{\alpha}_{\phantom{\alpha}\mu \nu}$, and the curvature tensor $R^{\alpha}_{\phantom{\alpha}\beta\mu \nu}$, defined respectively as \cite{Heisenberg:2023lru}:
\begin{equation}
    Q_{\alpha \mu \nu} = \nabla_{\alpha} g_{\mu \nu},
\end{equation}
\begin{equation}
    T^{\alpha}_{\phantom{\alpha}\mu \nu} = \Gamma^{\alpha}_{\phantom{\alpha}\nu \mu} - \Gamma^{\alpha}_{\phantom{\alpha}\mu \nu},
\end{equation}
\begin{equation}
    R^{\alpha}_{\phantom{\alpha}\beta\mu \nu} = \partial_\mu \Gamma^\alpha_{\phantom{\alpha}\beta\nu} - \partial_\nu \Gamma^\alpha_{\phantom{\alpha}\beta\mu} + \Gamma^\alpha_{\phantom{\alpha}\sigma\mu} \Gamma^\sigma_{\phantom{\alpha}\beta\nu} - \Gamma^\alpha_{\phantom{\alpha}\sigma\nu} \Gamma^\sigma_{\phantom{\alpha}\beta\mu}.
\end{equation}

In General Relativity (GR), Einstein adopted the Levi-Civita connection $\mathring{\Gamma}^{\alpha}_{\phantom{\alpha}\mu \nu}$, which assumes a vanishing non-metricity ($Q_{\alpha \mu \nu} = 0$) and vanishing torsion ($T^{\alpha}_{\phantom{\alpha}\mu \nu} = 0$). This connection is uniquely determined by the metric tensor as:
\begin{equation}
    \mathring{\Gamma}{^\alpha}_{\mu \nu} = \frac{1}{2} g^{\alpha \sigma} \left( \partial_\nu g_{\sigma \mu} + \partial_\mu g_{\sigma \nu} - \partial_\sigma g_{\mu \nu} \right).
\end{equation}
Consequently, the metric becomes the sole fundamental dynamical quantity in GR. However, one can make an alternative choice for the connection. In the Teleparallel Equivalent of General Relativity (TEGR), we assume a vanishing non-metricity ($Q_{\alpha \mu \nu} = 0$) and a vanishing curvature ($R^{\alpha}_{\phantom{\alpha}\beta\mu \nu} = 0$). In this framework, the fundamental dynamical variable is the tetrad field $e^a{}_\mu$, which is related to the spacetime metric $g_{\mu\nu}$ through
\begin{equation}
    e^a{}_\mu e^b{}_\nu g^{\mu\nu} = \eta^{ab}, \quad e^a{}_\mu e^b{}_\nu \eta_{ab} = g_{\mu\nu},
\end{equation}
where $\eta_{ab} = \mathrm{diag}(-1, 1, 1, 1)$ is the Minkowski metric of the tangent space. Geometrically, the tetrad defines an orthonormal basis for the tangent space at each spacetime point. Consequently, the general affine connection $\Gamma^\alpha{}_{\mu\nu}$ can be expressed as \cite{Maluf:2013gaa,2013tgif.book.....A}:
\begin{equation}\label{eq:conn}
    \Gamma^\alpha{}_{\mu\nu} = e_a{}^\alpha \partial_\nu e^a{}_\mu + e_a{}^\alpha \omega^a{}_{b\nu} e^b{}_\mu,
\end{equation}
where $\omega^a{}_{b\nu}$ is the spin connection corresponding to the tetrad. In the teleparallel framework, one can always select a specific frame (referred to as an inertial frame) such that the spin connection vanishes, i.e., $\omega^a{}_{b\mu} = 0$. Substituting this choice into Eq.~\eqref{eq:conn}, 
this leads to the Weitzenb\"{o}ck connection, defined as:\cite{Heisenberg:2023lru,2013tgif.book.....A,DeAndrade:2000sf}
\begin{equation}
    \Gamma{^\alpha}_{\mu\nu} = e{_a}^\alpha \partial_\nu e{^a}_\mu,
\end{equation}

The fundamental geometric object in this framework is the torsion scalar $T$ that can be defined as:
\begin{equation}
    T = \frac{1}{4} T^{\rho\mu\nu} T_{\rho\mu\nu} + \frac{1}{2} T^{\rho\mu\nu} T_{\nu\mu\rho} - T_{\rho}^{\phantom{\rho}\rho\mu} T^\nu_{\phantom{\nu}\nu\mu}.
\end{equation}
Furthermore, we define the contorsion tensor $K^\rho_{\phantom{\rho}\mu\nu}$ as follows:
\begin{equation}
K^\alpha_{\;\;\beta\gamma} = \frac{1}{2} \left(
T_\beta{}^\alpha{}_\gamma + T_\gamma{}^\alpha{}_\beta - T^\alpha_{\;\;\beta\gamma}
\right).
\end{equation}
It can be shown that the curvature scalar $\mathring{R}$ associated with the Levi-Civita connection differs from the torsion scalar $T$ associated with the Weitzenb\"{o}ck connection by a total derivative term:
\begin{equation}
    \mathring{R} = -T + 2\mathring\nabla_\mu ( T{^\lambda}_{\lambda}{^\mu}).
\end{equation}
Where $\mathring{\nabla}$ denotes the covariant derivative associated with the Levi-Civita connection.This relation ensures that an action linear in $T$ is equivalent to GR. To explore modifications to gravity, we generalize the action to a non-linear function $f(T)$:
\begin{equation}
    S = \int d^4x e \left[ \frac{-1}{2\kappa^2} f(T) + \mathcal{L}_m \right],
\end{equation}
where $e = \det(e{^a}_\mu)$ , $\kappa^2 = 8\pi G/c^4$ and $\mathcal{L}_{m}$ denote the Lagrangian scalar of matter.Varying this action with respect to the tetrad yields the field equations \cite{Li:2011wu}:
\begin{equation}
    f_T G_{\mu\nu} - \frac{1}{2} g_{\mu\nu} (T f_T - f) + S_{\mu\nu}^{\phantom{\mu\nu}\sigma} \partial_\sigma T f_{TT} = \kappa^2 \mathcal{T}_{\mu\nu},
    \label{fid_eq}
\end{equation}
where $f_T = \frac{df}{dT}$, $f_{TT}=\frac{d^2f}{dT^2}$, $G_{\mu\nu}$ is the Einstein tensor, and $S_{\rho}^{\phantom{\rho}\mu\nu}$ is the superpotential defined by:
\begin{equation}
    S_\rho^{\phantom{\rho}\mu\nu} = K^{\mu\nu}_{\phantom{\mu\nu}\rho} + \delta^\mu_\rho T^{\alpha\nu}_{\phantom{\alpha\nu}\alpha} - \delta^\nu_\rho T^{\alpha\mu}_{\phantom{\alpha\mu}\alpha}.
\end{equation}
$\mathcal{T}_{\mu\nu}$ is the energy-momentum tensor and is defined as :
\begin{equation}
    \mathcal{T}{_\nu} {^\mu}=-\frac{e{_a}^{\mu}}{e}\frac{\delta (e\mathcal{L}_m)}{\delta e{_a}^{\nu}}
\end{equation}
for a perfect fluid is given by:
\begin{equation}
    \mathcal{T}_{\mu\nu} = \rho u_\mu u_\nu + p(g_{\mu\nu} + u_\mu u_\nu),
\end{equation}
satisfying the conservation law $\mathring\nabla_\mu \mathcal{T}^{\mu\nu} = 0$.
\subsection{\texorpdfstring{$f(T)$}{} Cosmology}
In a flat Friedmann-Lema\^{i}tre-Robertson-Walker (FLRW) background, the line element is defined as: \begin{equation}
    ds^2 = -dt^2 + a^2(t) \delta_{ij} dx^i dx^j \label{eq:lin_ele}
\end{equation} ,where $a(t)$ is the scale factor .This line element corresponds to the diagonal tetrad $e{^a}_\mu = \text{diag}(1, a, a, a)$. Substituting this into the field equations \eqref{fid_eq}, we obtain the modified Friedmann equations:
\begin{equation}
    3H^2 = \kappa^2 (\rho_m + \rho_r) + \frac{1}{2} (-2T f_T + f + T),
    \label{eq:fir_frid}
\end{equation}
\begin{equation}
    \dot{H} = -\frac{\kappa^2}{2} \left( \frac{\rho_m + p_m + \rho_r + p_r}{f_T + 2T f_{TT}} \right),
\end{equation}
where $H = \dot{a}/a$ and $T = 6H^2$. In the late-time universe (where we can neglect $\rho_r$ and $p_r$), these equations can be rewritten by defining an effective torsion density $\rho_T$ and pressure $p_T$:
\begin{equation}
    3H^2 = \kappa^2 (\rho_m + \rho_T), \quad \dot{H} = -\frac{\kappa^2}{2} (\rho_m + p_m + \rho_T + p_T),
\end{equation}
where:
\begin{equation}
    \rho_T = \frac{1}{2\kappa^2} (-2T f_T + T + f), \quad p_T = \frac{1}{2\kappa^2} \left( \frac{-f + T f_T - 2T^2 f_{TT}}{f_T + 2T f_{TT}} \right). \label{eq:rhop}
\end{equation}
The torsion equation of state $w_T = p_T/\rho_T$ and the effective equation of state 
\begin{equation}
w_{\text{eff}} = -1 - 2\dot{H}/3H^2,
\end{equation}
characterize the expansion dynamics. Finally, each component satisfies the continuity equation:
\begin{equation}
    \dot{\rho}_i + 3H(1+w_i)\rho_i = 0.\label{eq:cnt}
\end{equation}

To study the growth of structures and the Cosmic Microwave Background (CMB) anisotropies, we consider linear scalar perturbations. In the Newtonian gauge, the perturbed Friedmann-Lema\^{i}tre-Robertson-Walker (FLRW) metric is given by:
\begin{equation}
    ds^2 = a^2(\tau) \left[ -(1 + 2\psi) d\tau^2 + (1 - 2 \phi) \delta_{ij} dx^i dx^j \right],
\end{equation}
where $\tau$ is the conformal time, and $\phi$ and $\psi$ are the gravitational potentials. A general perturbed tetrad that reproduces the above metric \cite{Golovnev:2018wbh} is:
\begin{equation}
    e^0_0 = a(1 + \psi), \quad e^0_i = a \partial_i \xi, \quad e^a_0 = a \partial^a \xi, \quad e^a_j = a \left[ (1 - \phi) \delta^a_j + \epsilon^a_{\phantom{a}jk} \partial^k S \right],
\end{equation}
where $\xi$ represents the scalar part of the local Lorentz boost and $S$ corresponds to the scalar part of spatial rotations. Unlike General Relativity, $f(T)$ gravity is not invariant under local Lorentz transformations, which introduces these additional degrees of freedom. While $S$ remains unconstrained and does not enter the field equations at the linear level, the variable $\xi$ is determined by the antisymmetric part of the field equations.

In Fourier space, the mixed component of the antisymmetric field equations yields a constraint for $\xi$:
\begin{equation}
    k^2 \xi = 3 \left( \phi' + \mathcal{H}\psi - \frac{\mathcal{H}' - \mathcal{H}^2}{\mathcal{H}} \phi \right),
\end{equation}
where a prime denotes a derivative with respect to conformal time and $\mathcal{H} = a'/a$ is the conformal Hubble parameter. 

The symmetric part of the field equations provides the modified Poisson-like equations. The first relates the two gravitational potentials:
\begin{equation} \label{eq:slib}
    \psi = \phi - 12\pi Q G \left(\frac{a}{k}\right)^2 (\rho + p)\sigma - \mathcal{H} \Xi \xi,
\end{equation}
where we have defined the following auxiliary functions:
\begin{equation}
    Q = \frac{1}{f_T}, \quad \Xi = -12(\mathcal{H}' - \mathcal{H}^2) Q f_{TT}.
\end{equation}
The evolution of the curvature potential $\phi$ is obtained from the mixed temporal-spatial component of the symmetric field equations:
\begin{equation} \label{per2}
    \phi' = 4\pi Q G \left(\frac{a}{k}\right)^2 (\rho + p)\theta - \mathcal{H}\psi - \frac{\mathcal{H}\Xi}{a^2}\phi,
\end{equation}
where $\theta$ and $\sigma$ are the velocity divergence and the anisotropic stress of the fluid, respectively. The remaining metric components is the tensor perturbation $h_{ij}$, which is obtained from
\begin{equation}
    h''_{ij}+\mathcal{H}\left(2+\frac{\Xi}{a^2}\right)h'_{ij}+k^2h_{ij}=0.
\end{equation}
Finally, the conservation of the energy-momentum tensor leads to the perturbed continuity and Euler equations for each fluid species:
\begin{align}
    \delta' &= -(1+w) (\theta - 3 \phi') + 3 \mathcal{H} \left( w - c_s^2 \right) \delta, \\
    \theta' &= -\mathcal{H} (1 - 3 w) \theta - \frac{w'}{1 + w} \theta + \frac{c_s^2}{1 + w} k^2 \delta - k^2 \sigma + k^2 \psi,
\end{align}
where $\delta = \delta \rho / \rho$ is the density contrast and $c_s^2 = \delta p / \delta \rho$ is the adiabatic sound speed.These equations provide a complete theoretical framework, allowing $f(T)$ gravity to be rigorously compared against all cosmological data sets, encompassing both background expansion and growth of perturbations \cite{Hashim:2020sez,Hashim:2021pkq,Escamilla-Rivera:2024sae}.

\section{New Exponential Infrared Model II}\label{sec:model}
Modified gravity models \cite{Nojiri:2017ncd,CANTATA:2021asi} have emerged as one of the most promising approaches for addressing several outstanding problems in modern cosmology, including the late-time accelerated expansion of the Universe \cite{SupernovaSearchTeam:1998fmf,SupernovaCosmologyProject:1998vns} and the Hubble tension \cite{Cai:2026swf}. Among these theories, $f(T)$ teleparallel gravity \cite{Bahamonde:2021gfp} provides a particularly attractive framework, as it represents a minimal extension of General Relativity while preserving second-order field equations, in contrast to many other modified gravity theories.

Within this framework, an exponential infrared $f(T)$ model was first proposed as an alternative to the $\Lambda$CDM model in Ref.~\cite{Awad:2017yod}. Notably, this model does not introduce additional free cosmological parameters beyond those of the baseline $\Lambda$CDM model and does not produce $\Lambda\text{CDM}$ as a special case as many models did ~\cite{Nesseris:2013jea} :
\begin{equation}
f(T)=Te^{\beta\frac{T_0}{T}},
\end{equation}
where $T_0$ is the present value of the torsion scalar, and $\beta$ is the model parameter, which can be expressed in terms of the present torsion density parameter through the Lambert function $W$. This function admits two real branches, $W_0$ and $W_{-1}$, leading to two distinct solutions for $\beta$. The principal $W_0$ branch was subsequently tested against both background observations and cosmological perturbation data in Refs.~\cite{Hashim:2020sez,Hashim:2021pkq}. The secondary branch associated with the $W_{-1}$ solution was later identified and studied, revealing a qualitatively different cosmological evolution characterized by an effective dark-energy density that evolves from negative to positive values at late times \cite{Akarsu:2024nas}. This branch was also confronted with cosmological observations, providing constraints on its observational viability \cite{Hashim:2026yoy}.

This model was subsequently generalized by introducing an additional parameter $b$ in Ref.~\cite{Santos:2022atq}:
\begin{equation}
f(T)=Te^{\beta\left(\frac{T_0}{T}\right)^b}.
\end{equation}
It was first investigated using background cosmological observations only, and in Ref.~\cite{Hashim:2026M2}, the analysis was extended to include perturbation data, providing the first complete cosmological constraints on the model. More recently, a new class of infrared exponential $f(T)$ models was introduced in Ref.~\cite{Hashim:2026M3}:
\begin{equation}
f(T)=\left(T+\alpha\sqrt{T_0T}\right)e^{\beta\frac{T_0}{T}}, \label{eq:new_model}
\end{equation}
with $\alpha$ as a free parameter, providing a broader framework that encompasses some previously proposed models as special cases. This model is motivated by the fact that the term $\alpha\sqrt{T_0T}$ contributes only as a boundary term (i.e., a total derivative) at the background level for any constant value of $\alpha$. Consequently, in the early-Universe limit ($T\gg T_0$), the exponential factor approaches unity, and Eq.~\eqref{eq:new_model} asymptotically reduces to
\begin{equation}
f(T)\sim T+\alpha\sqrt{T_0T}+\beta T_0,
\end{equation}
which is dynamically equivalent to General Relativity with a cosmological constant at the background level. In the late-time Universe, the exponential factor becomes non-negligible, driving a cosmological evolution distinct from that of $\Lambda$CDM.This makes it a viable model to test against observational data.

In our previous work \cite{Bedair:2026M4}, we investigated a particular case of the New Exponential Infrared $f(T)$ teleparallel gravity model,
\begin{equation}
f(T)=\left(T-\sqrt{TT_0}\right)\exp\left(\beta\frac{T_0}{T}\right).
\label{eq:old_model}
\end{equation}
The model in Eq.~\eqref{eq:old_model} corresponds to the choice $\alpha=-1$ in Eq.~\eqref{eq:new_model}, a choice motivated by the results of Ref.~\cite{Hashim:2026M3}, where the combined cosmological datasets consistently favored values of $\alpha\to-1$. In Ref.~\cite{Bedair:2026M4}, we showed that this model behaves as $\Lambda_s$CDM in the early Universe (with $\Lambda_s<0$) \cite{Akarsu:2021fol} and gives rise to an effective dark-energy sector that undergoes a sign transition from negative to positive values at $z_{\text{tr}}\sim1.6$, without introducing any additional free cosmological parameters beyond the standard baseline set. Although that choice yielded a viable sign-changing dark-energy sector, it represents a mathematically degenerate limit of the new exponential infrared model \eqref{eq:new_model}.

In the present work, we investigate a representative non-degenerate case corresponding to $\alpha=+1$,
\begin{equation}
f(T)=\left(T+\sqrt{T_0T}\right)\exp\left(\beta\frac{T_0}{T}\right),
\label{eq:model}
\end{equation}
which leads to a fundamentally different cosmological evolution. The origin of this difference lies in the equation that determines the model parameter $\beta$. In the general case, $\beta$ is obtained from an equation involving the Lambert function $W$, as will be shown below, whereas the $\alpha=-1$ model corresponds to a highly restrictive limiting case in which the Lambert $W$ structure disappears and the solution reduces to a logarithmic expression as shown in Ref.~\cite{Bedair:2026M4}.

By substituting Eq.~\eqref{eq:model} into the first Friedmann equation \eqref{eq:fir_frid} and normalizing by $3H_0^2$, we obtain \cite{Nesseris:2013jea}:
\begin{equation}
    E^2(z) = \Omega_{m0} (1+z)^3 + \Omega_{r0} (1+z)^4 + \Omega_{T0} \, y(E, \beta), \label{eq:nrmlzd}
\end{equation}
where the dimensionless torsion density contribution is defined as:
\begin{equation}
    y(E, \beta) = \frac{1}{T_0 \Omega_{T0}} \left[ T + f(T) - 2T f_T(T) \right],
\end{equation}
and 
\begin{equation}
    \Omega_{T0}=\frac{\kappa^2 \rho_T(0)}{3H_0^2}.
\end{equation}
For the model under consideration \eqref{eq:model}, the function $y(E, \beta)$ takes the form:
\begin{equation}
    y(E, \beta) = \frac{1}{\Omega_{T0}} \left[ E^2 + \exp\left(\frac{\beta}{E^2}\right) \left( -E^2 + 2\beta + \frac{2\beta}{E} \right) \right]. \label{eq:y_E}
\end{equation}
Applying the normalization condition at the present epoch, $y(1,\beta) = 1$, we derive a relation for $\beta$:
\begin{equation}
    \left( \beta - \frac{1}{4} \right) \exp\left( \beta - \frac{1}{4} \right) = \frac{\Omega_{T0} - 1}{4} \exp\left(-\frac{1}{4}\right).
\end{equation}
The solution for $\beta$ is expressed in terms of the Lambert function$W$  as:
\begin{equation}
    \beta = \frac{1}{4} + W\left( \frac{\Omega_{T0} - 1}{4} \exp\left[-\frac{1}{4}\right] \right). \label{eq:beta_lambert}
\end{equation}
From the field equation at $z = 0$, it follows that $\Omega_{T0}=1-\Omega_{r0}-\Omega_{m0}$. Given that $0 < \Omega_{T0} < 1$ for a viable late-time cosmology, the argument of the Lambert function lies in the interval $(-1/e, 0)$, where the function is double-valued. This gives rise to two distinct branches for the model parameter:
\begin{equation}
    \beta^{I} = \frac{1}{4} + W_0\left( \frac{\Omega_{T0} - 1}{4} \exp\left[-\frac{1}{4}\right] \right),\label{eq:prnc}
\end{equation}
\begin{equation}
    \beta^{II} = \frac{1}{4} + W_{-1}\left( \frac{\Omega_{T0} - 1}{4} \exp\left[-\frac{1}{4}\right] \right).\label{eq:scnd}
\end{equation}
These two values represent two distinct cosmological models. For a standard baseline value of $\Omega_{T0} = 0.7$, the calculated parameters from Eqs.~\eqref{eq:prnc} and \eqref{eq:scnd} are:
\begin{equation}
    \beta^{I} \approx 0.1818, \quad \beta^{II} \approx -4.
\end{equation}
For $0<\Omega_{T0}<1$, we always obtain $\beta^{I}>0$ and $\beta^{II}<0$. The impact of these two distinct values of $\beta$ on the cosmic evolution can be further understood by examining the asymptotic behavior of Eq.~\eqref{eq:y_E}. In the early-time limit ($E \gg 1$), the function $y(E, \beta)$ simplifies to:
\begin{equation}
    y(E, \beta) \approx \frac{\beta}{\Omega_{T0}}.
\end{equation}
For the second branch, $\beta = \beta^{II} < 0$, the effective dark energy density starts with a negative value in the early Universe, eventually undergoing a transition to positive values at late times. This behavior is qualitatively similar to the model presented in Ref.~\cite{Bedair:2026M4}, yet the significantly larger magnitude of $\beta^{II}$ here leads to a quantitatively distinct expansion history.

\begin{figure}[t!]
    \centering
    \begin{subfigure}{0.48\textwidth} 
        \centering
        \includegraphics[width=\linewidth]{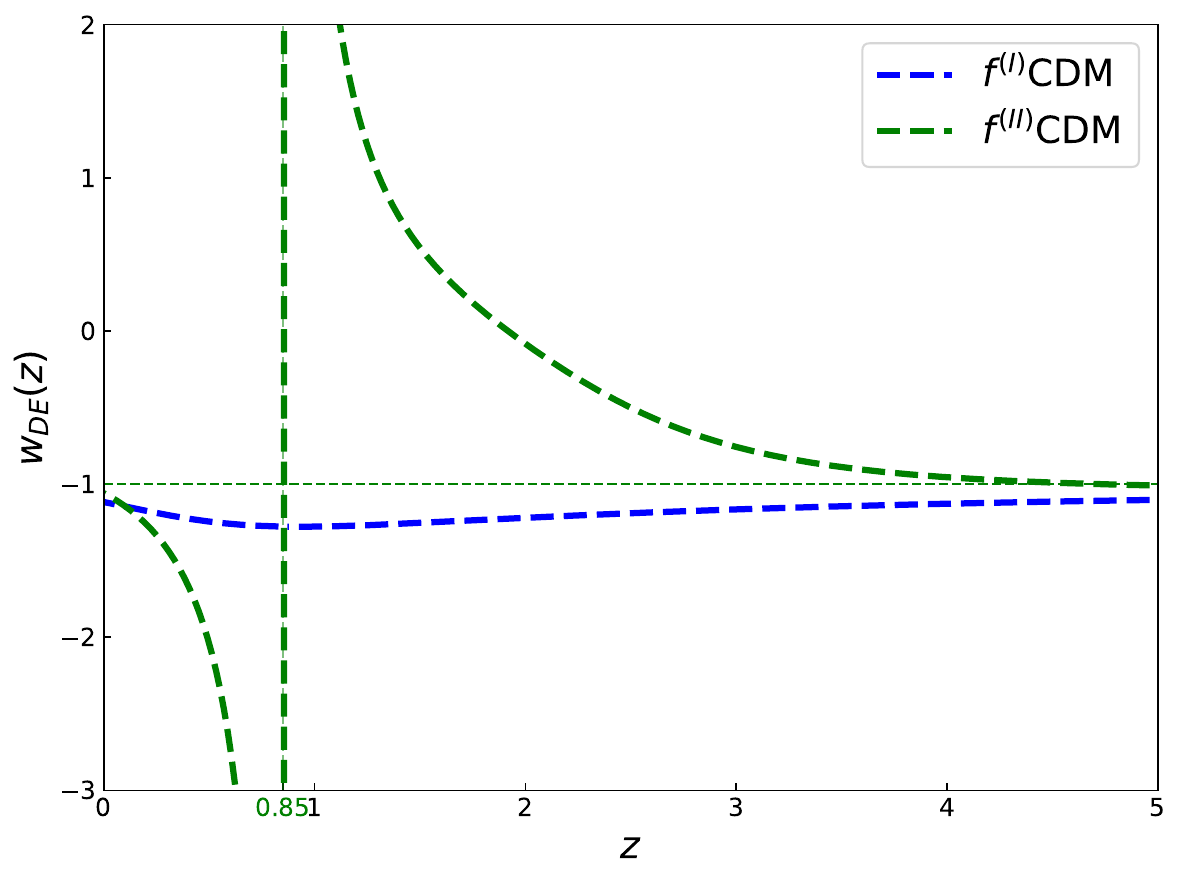}
        \caption{Evolution of the EoS parameter $w_T$ \eqref{eq:EOS}}
        \label{fig:DEOS}
    \end{subfigure}
    \hfill
    \begin{subfigure}{0.48\textwidth}
        \centering
        \includegraphics[width=\linewidth]{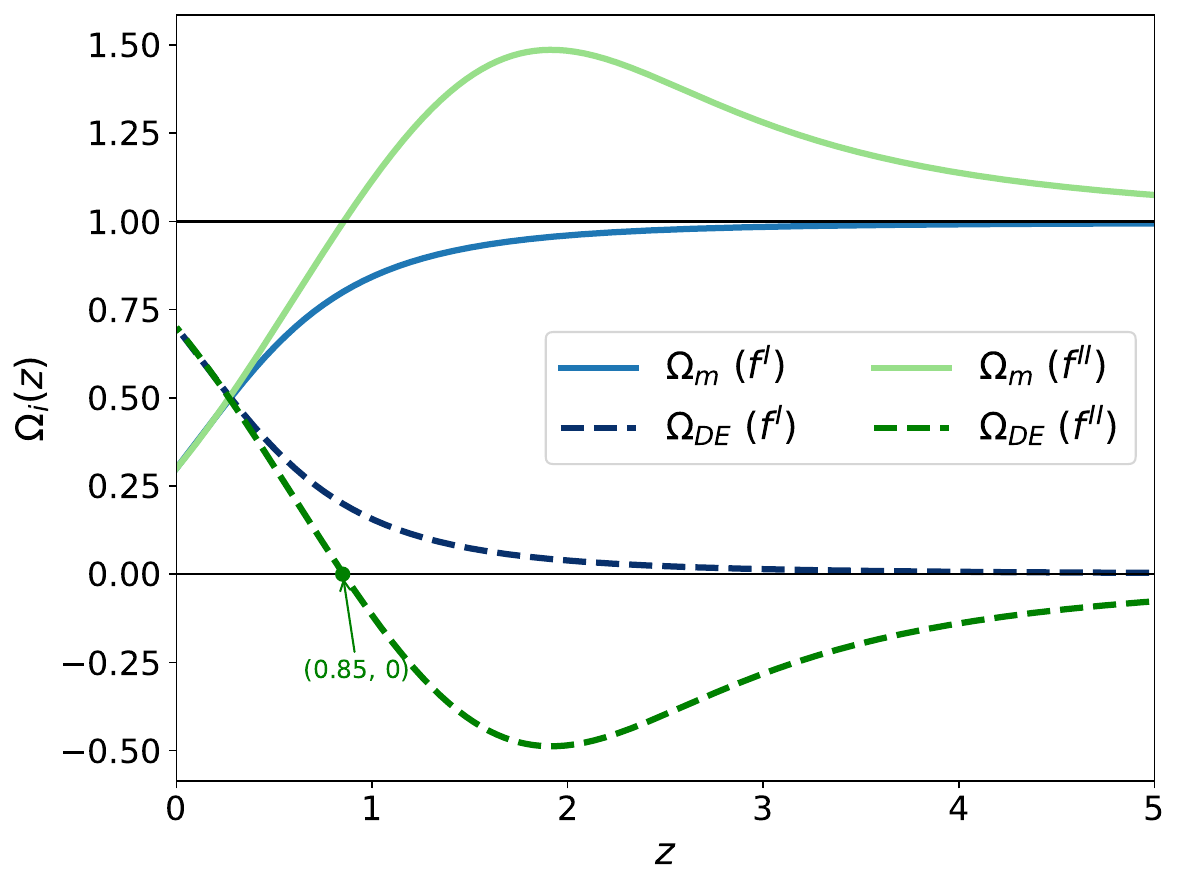}
        \caption{Evolution of density parameters \eqref{eq:Pdnsm}, \eqref{eq:Pdnstt}}
        \label{fig:dnstprmtr}
    \end{subfigure}
    \caption{Assuming the parameter values $\Omega_{m0}=0.3$ and $H_0=70\text{ km s}^{-1}\text{Mpc}^{-1}$. (\subref{fig:DEOS}) Evolution of the effective torsion equation-of-state parameter, $w_T$, for the model \eqref{eq:model}. The blue and green dashed curves correspond to the $f^{(I)}$ and $f^{(II)}$ branches, obtained from the principal and secondary Lambert branches of Eq.~\eqref{eq:beta_lambert}, respectively. The vertical green dashed line marks the singularity of $w_T$ at $z\simeq0.85$, coinciding with the sign change of the effective torsion density, $\rho_T$. (\subref{fig:dnstprmtr}) Evolution of the density parameters for the two branches. In the $f^{(I)}$ model, $\Omega_m<1$ throughout the evolution owing to the positive torsion density ($\Omega_T>0$). In contrast, the $f^{(II)}$ branch exhibits $\Omega_m>1$ for $z>0.85$, reflecting the negative torsion density ($\Omega_T<0$), with the transition occurring at $z=0.85$.}
    \label{fig:bckgrnd}
\end{figure}

Conversely, for the first branch where $\beta = \beta^{I} > 0$, the model exhibits a positive effective dark energy density throughout the entire cosmic evolution. Notably, this branch shows a persistent phantom-like behavior ($w_{T} < -1$) in the present stages of the Universe, as illustrated in Fig.~\ref{fig:DEOS}.

For the remainder of this work, we shall denote the model defined by Eq.~\eqref{eq:model} with the parameter $\beta^{I}$ as the $f^{(I)}$ model, and the case with $\beta^{II}$ as the $f^{(II)}$ model.

Substituting the functional form \eqref{eq:model} into the definitions of the effective torsion density and pressure \eqref{eq:rhop}, we obtain:
\begin{equation} \label{eq:rho}
    \rho_T = \frac{3H_0^2}{\kappa^2} \left[ E^2 + \exp\left(\frac{\beta}{E^2}\right) \left( -E^2 + 2\beta + \frac{2\beta}{E} \right) \right],
\end{equation}
\begin{equation} \label{eq:press}
    p_T = -\frac{3H_0^2\beta E^2 }{\kappa^2} \left( \frac{E^3 + 3E^2 + 2\beta E + 2\beta}{E^5 - \beta E^3 + \beta E^2 + 2\beta E + 2\beta^2} \right).
\end{equation}
Consequently, the Equation of State (EoS) parameter for the torsion fluid, $w_T = p_T / \rho_T$, is derived as:
\begin{equation}
    w_T = \frac{-\beta \left( E^3 + 3E^2 + 2\beta E + 2\beta \right)}{\left( E^5 - \beta E^3 + \beta E^2 + 2\beta E + 2\beta^2 \right) \left[ 1 + \exp(\beta/E^2) \left( -1 + 2\beta/E^2 + 2\beta/E^3 \right) \right]}.\label{eq:EOS}
\end{equation}

In Fig.~\ref{fig:DEOS}, we illustrate the evolution of the EoS parameter $w_T$ for the two branches of the model, denoted as $f^{(I)}$ and $f^{(II)}$, assuming the baseline parameters $\Omega_{m0} = 0.3$ and $H_0 = 70\text{ km s}^{-1}\text{Mpc}^{-1}$.

As depicted in the figure, the $f^{(I)}$ model ($\beta^I > 0$) originates with a de Sitter-like behavior ($w_T \to -1$) in the early Universe, subsequently evolving into a phantom-crossing regime ($w_T < -1$) at late times and through the present epoch. In contrast, the $f^{(II)}$ model ($\beta^{II}<0$) exhibits a more complex evolution. It starts with $\rho_T<0$ and an equation-of-state parameter close to $w_T\sim -1$. As the Universe evolves, the model enters a quintessence-like regime ($w_T>-1$), before encountering a mathematical singularity at $z\sim0.85$. This divergence in the EoS parameter corresponds to the null-point of the torsion density ($\rho_T = 0$), where the effective energy density undergoes a transition from negative to positive values. Following this transition, the $f^{(II)}$ model evolves as an effective phantom fluid toward the current era.

Models in which the effective dark-energy density evolves from a negative value at relatively high redshifts to a positive value at late times have recently been discussed in Refs.~\cite{Akarsu:2021fol,Menci:2024rbq,Gomez-Valent:2024ejh,Soriano:2025gxd}. In those studies, an additional independent parameter, $z_{\rm tr}$, was typically introduced to control the redshift at which the effective dark-energy density changes sign, and the transition was not necessarily smooth. In contrast, in the $f^{(II)}$CDM model, the sign change occurs smoothly, with $\rho_T$ continuously evolving from negative to positive values , without introducing any extra parameters. This type of sign-changing dark-energy behavior was originally proposed as a possible mechanism to address the tension between the Ly$\alpha$ BAO measurements and the $\Lambda$CDM model constrained by Planck data \cite{BOSS:2014hhw}. In particular, it was found that models with a negative effective dark-energy density, $\rho_{\rm eff}<0$, at $z\gtrsim1.6$ can alleviate this tension.

We define the density parameters as $\Omega_i = \rho_i / \rho_c$, where $\rho_c = 3H^2 / \kappa^2$ represents the critical density. Consequently, the matter density parameter is expressed as:
\begin{equation}
    \Omega_m = \frac{\kappa^2 \rho_m}{3H^2} = \frac{\Omega_{m0}(1+z)^3}{E^2}. \label{eq:Pdnsm}
\end{equation}
Similarly, the torsion dark energy density parameter is obtained as:
\begin{equation}
    \Omega_T = \frac{\kappa^2 \rho_T}{3H^2} = \frac{\Omega_{T0} \, y(E(z), \beta)}{E(z)^2}. \label{eq:Pdnstt}
\end{equation}

Neglecting the radiation contribution at late times ($z < 5$), solving Eq.~\eqref{eq:nrmlzd} for the redshift $z$ yields:
\begin{equation}
    z(E) = \left( \frac{E^2 - \Omega_{T0} \, y(E, \beta)}{\Omega_{m0}} \right)^{\frac{1}{3}} - 1. \label{eq:z}
\end{equation}

Using the inverse relation of Eq.~\eqref{eq:z} in Eqs.~\eqref{eq:Pdnsm} and \eqref{eq:Pdnstt}, we obtain the evolution of the density parameters in terms of the redshift $z$. In Fig.~\ref{fig:dnstprmtr}, we plot the density parameters for the two models $f^{(I)}$ and $f^{(II)}$ within the interval $0 < z < 5$.

\section{$\text{Phase Portrait}$}\label{sec:phase}
From the general dynamical equation of the $f(T)$ \text{model} (see \cite{Awad:2017yod}), we have the following.
\begin{equation}
    \dot{H}
    = 
    3(1+\omega)\,
    \frac{f - H f_H}{f_{HH}}.
\end{equation}

Where $f_H=\frac{df(6H^2)}{dH}$,$f_{HH}=\frac{d^2f(6H^2)}{dH^2}$. For functional form \eqref{eq:model}, we obtain the following:
\begin{equation}
\dot{H}
= 
-\frac{3(1+\omega)}{2}\,
\frac{
H^4\!\left(H^3 - 2\beta H H_0^2 - 2\beta H_0^3\right)
}{
H^5 - \beta H_0^2 H^3 + \beta H_0^3 H^2 + 2\beta^{\,2} H_0^4 H + 2\beta^{\,2} H_0^5
}.\label{eq:hdot}
\end{equation}
Eq. \eqref{eq:hdot} can be rewritten in an alternative form as:
\begin{equation}
 \dot{H} = -\frac{3(1+\omega)H^2_0}{2} \frac{ E^4 \left(E^3 - 2\beta E - 2\beta\right)}{E^5 - \beta E^3 + \beta E^2 + 2\beta^{2} E + 2\beta^{2}}, \label{eq:hdotmod}
\end{equation}
Eq.\eqref{eq:hdotmod} shows that varying $H_0$ does not qualitatively alter the behavior of the fixed points; that is, the number and stability of these points remain invariant, while their positions change only quantitatively. Consequently, a detailed analysis performed at a specific value of $H_0$ can be generalized across the whole physical parameter space. Similarly, for the preferred cosmological range of the density parameter $\Omega_{m0}$, the  behavior of the system remains qualitatively consistent, undergoing only quantitative shifts. Thus, the following dynamical study maintains a sufficient degree of generality.

We  analyze the phase portrait separately for the principal branch $f^{(I)}$ ($\beta=\beta^I$) \eqref{eq:prnc} and for the secondary branch $f^{(II)}$ $(\beta=\beta^{II})$  \eqref{eq:scnd}.

The fixed points of system \eqref{eq:hdot} satisfy
\begin{equation}
H = 0, \qquad H = H_0 E,
\end{equation}
where $E$ satisfy the following
\begin{equation}
E^3 - 2\beta E - 2\beta = 0.\label{eq:fixed}
\end{equation}
For both branches, Eq.~\eqref{eq:fixed} admits a single real de Sitter solution, while the denominator of Eq.~\eqref{eq:hdot} possesses only one real root, corresponding to a Type II sudden singularity and it always lies in the negative $H$ region.
Fig.\ref{fig:phase} shows the phase portraits of the two branches of the model, $f^{(I)}$ and $f^{(II)}$, for $\Omega_{T0}=0.7$ and $H_0=70$ km s$^{-1}$ Mpc$^{-1}$.

\subsection*{$\text{Primary branch}$ \( f^{(I)} \) (blue-curve in Figure \ref{fig:phase})}

In this branch there is exactly one de Sitter fixed point ($H^\ast > 0$), while the zero of the denominator ($H_s<0$) where $\dot{H}$ has a singurality (diverges).  
Thus:
- For $H > H^\ast$, the universe begins at a big–bang singularity and evolves toward the de Sitter point $H^\ast$ (represented by
blue shaded circle in Figure \ref{fig:phase}) in infinite time.
- For $0 < H < H^\ast$, the universe approaches $H^\ast$ from a phantom region ($\dot{H} > 0$), making the de Sitter fixed point ($H^\ast$) stable.
- For $H < H_{\rm s} < 0$, the universe evolves toward a big–crunch singularity in finite time.
- For $H_{\rm s} < H < 0$, the system approaches Minkowski fixed-point ($H=0$)(represented by
gry shaded circle in Figure \ref{fig:phase}) from a phantom region and reaches it in infinite time, making Minkowski semi–stable.
  
Although the model admits a Type II (sudden) singularity at $H=H_{\rm s}$,(represented by blue vertical dashed line in Figure \ref{fig:phase}) it remains physically inaccessible. Consequently, the universe will not encounter this singularity at any point throughout its entire evolution, ensuring a smooth and non-singular cosmic history.

\begin{figure}[t!]
    \centering
    \includegraphics[width=.7\linewidth]{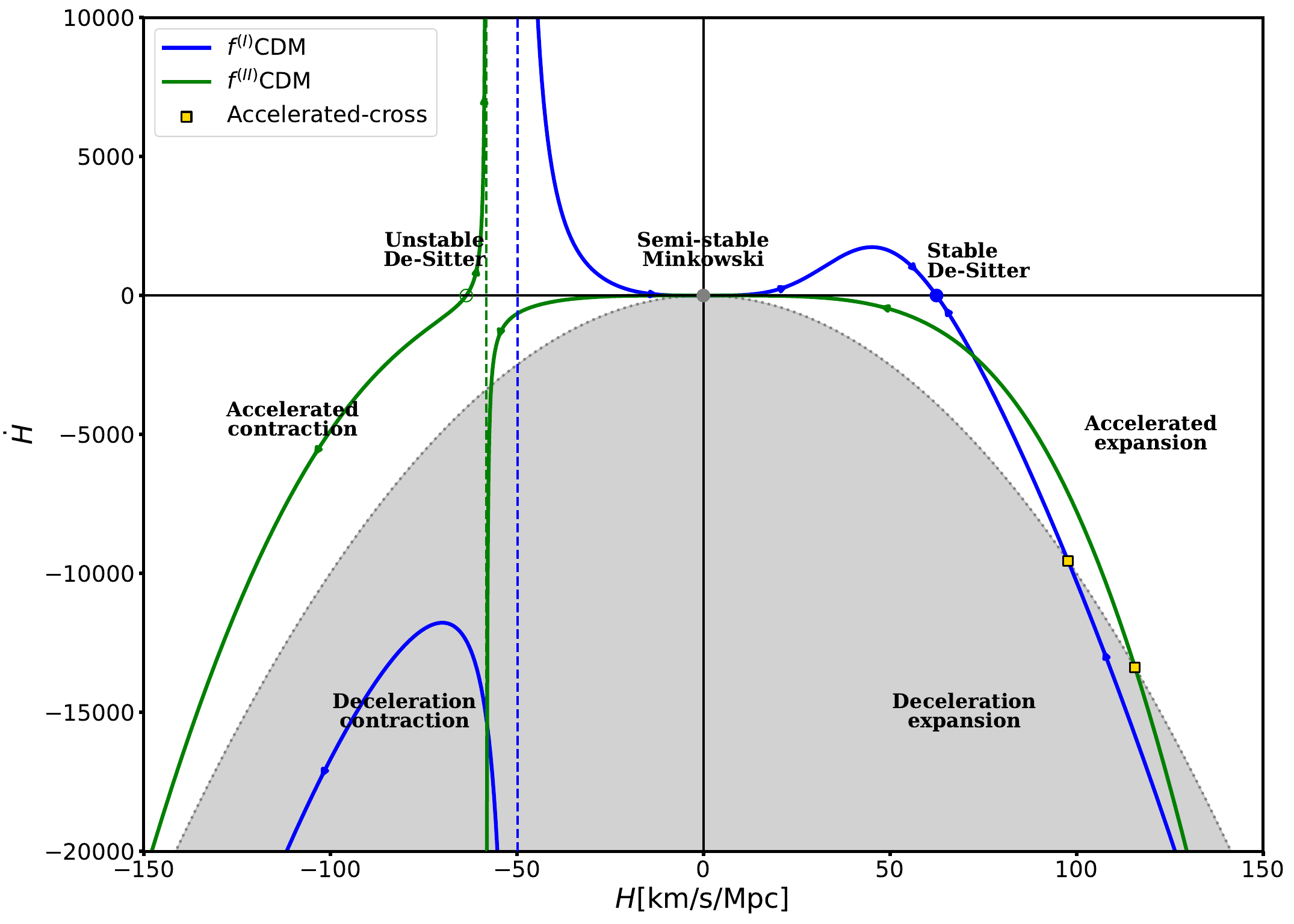}
    \caption{Phase portrait of the $f^{(I)}$ and $f^{(II)}$ models for the parameter set $\{\Omega_{T0}=0.7, H_0=70\}$. The blue (green) solid curve corresponds to the evolution of the $f^{(I)}$ ($f^{(II)}$) branch. The filled blue circle and the open green circle denote the stable and unstable de Sitter fixed points, respectively, while the corresponding Type II sudden singularities are indicated by the blue and green dashed vertical lines. The dotted curve represents the zero-acceleration boundary, $\dot{H}=-H^2$ (equivalently $q=0$), separating the accelerating and decelerating expansion regimes. The yellow square marks the transition from deceleration to acceleration, and the gray circle denotes the Minkowski fixed point at $H=0$.
}
    \label{fig:phase}
\end{figure}

\subsection*{$\text{Secondary branch}$ \( f^{(II)} \) (green-curve in Figure \ref{fig:phase})}

Here the ordering of the singularity and fixed point as following:  
\[
H^\ast < H_{\rm s} < 0.
\]

Then:
- For $H > 0$, the universe evolves from an initial singularity toward the Minkowski point in infinite time.
- For $H < H^\ast$, the universe evolves toward a big–crunch in finite time.
- For $H^\ast < H < H_{\rm s}$, the system flows towards a sudden singularity at $H_{\rm s}$ in finite time and becomes trapped there.
- For $H_{\rm s} < H < 0$, the system also evolves toward a sudden singularity in finite time. For $H < H^\ast < 0$, the universe evolves towards a big–crunch singularity in finite time.

Thus Minkowski is semi–stable, the de Sitter point is unstable (represented by
the green Hollow  circle in Figure \ref{fig:phase}), and the universe inevitably encounters a finite–time singularity (represented by green vertical dashed line in Figure \ref{fig:phase}) in this branch when starting from at some point in $\left(H^\ast,0\right)$.

\subsection{Redshift of the Deceleration–Acceleration Transition}\label{redshift}

The transition from deceleration to acceleration expansion occurs when the deceleration parameter changes sign, equivalently when
\begin{equation}
\dot{H} = -H_{acc}^2.
\end{equation}
Here $H_{\rm acc}$ denotes the Hubble function at the transition from the deceleration to the acceleration expansion. Substituting this into the dynamical equation \eqref{eq:hdot} gives
\begin{equation}
-E_{acc}^2 
= 
-\frac{3(1+\omega)}{2}\,
\frac{
E_{acc}^4\!\left(E_{acc}^3 - 2\beta E_{acc}  - 2\beta \right)
}{
E_{acc}^5 - \beta_2  E_{acc}^3 + \beta_2  E_{acc}^2 + 2\beta_2^{\,2}  E_{acc} + 2\beta^{\,2} 
}.
\end{equation}
where $E_{acc} = \frac{H_{acc}}{H_0}$.neglecting the radiation contribution, which is negligible at the late-time so we have $\omega=0$. For a given $\Omega_{T0}$, we can solve this equation after substituting $\beta^{I}$ from Eq.~\eqref{eq:prnc} to obtain $E_{acc}$ for the $f^{(I)}$ model, and similarly for the $f^{(II)}$ model after substituting $\beta^{II}$ from Eq.~\eqref{eq:scnd}.

To compute $z_{cc}$ (the redshift at the deceleration-acceleration transition),  using the Friedmann equation gives:
\begin{equation}
z_{acc}
=
\left(
\frac{E_{acc}^2 - (1-\Omega_{m0})\, y(E_{acc})}{\Omega_{m0}}
\right)^{1/3}
- 1.
\end{equation}

Using $\Omega_{m0}=0.3$ we obtain $(z_{\rm acc}=0.68,H_{\rm acc}=103.36 \text{ km s}^{-1} \text{ Mpc}^{-1})$ for $f^{(I)}\text{model}$ and $(z_{\rm acc}=1.28,H_{\rm acc}=122.306 \text{ km s}^{-1} \text{ Mpc}^{-1})$ for $f^{(II)}\text{model}$.

\section{Data and priors}\label{sec:data}
We utilized a modified version of the \texttt{CLASS}\protect\footnotemark code \cite{2011JCAP...07..034B} to compute the cosmological observables required for our data analysis. The parameter estimation was performed using the \texttt{Cobaya} framework \cite{Lewis:2019xzd}, which generated MCMC chains with $R-1<0.02$ and exploring the parameter space. Finally, the \texttt{GetDist} package \cite{Torrado:2020dgo} was employed to process the resulting chains, generate 1D posterior distributions and 2D contour plots, and estimate the best-fit values along with the associated variances for each cosmological parameter.  Table~\ref{tab:priors} shows the priors used for each parameter.
\footnotetext{
This code is publicly available at \href{https://github.com/mwhashim/class_tmg}{\faGithub} \url{https://github.com/mwhashim/class_tmg},
} 
In this work, we use several observational datasets to constrain our cosmological model. Each dataset is described below, including the likelihoods used in our analysis.
\subsection*{Type Ia Supernovae (SN~Ia)}
Supernova Type Ia serve as standardizable candles through the measurement of the peak apparent magnitude $m_{B}$, which allows us to infer the luminosity distance and therefore the distance modulus. The supernova dataset provides the necessary light-curve information to compute the relative distance from $m_{B}$, which we use directly in our analysis. The datasets and their corresponding likelihoods in Cobaya are:

     \textbf{Pantheon+} \cite{Brout:2022vxf}: An updated compilation containing approximately 1700 SN~Ia observations, making it one of the largest and most homogeneous SN datasets. Pantheon+ spans a wide redshift range ($0.01 < z < 2.6$). \\
    \textit{Likelihood used:} \texttt{sn.Pantheon\_Plus}
    
    \textbf{Union3} \cite{Rubin:2023jdq}: The latest release of the Union SN compilation series containing more than 2000 calibrated SN~Ia observations. \\
    \textit{Likelihood used:} \texttt{sn.Union3}
    
   \textbf{DES-SN} \cite{DES:2024jxu}: The five-year Dark Energy Survey supernova sample, providing a high-quality set of approximately 1600 SN~Ia. cover  the redshift range $0.1 < z < 1.13$. with  external low-redshift sample consisting of 194 SNe Ia spanning $0.025<z<0.1$ \\
    \textit{Likelihood used:} \texttt{sn.desy5}

\subsection*{Cosmic Microwave Background (CMB)}
Planck data represent measurements of the temperature anisotropies and polarization of the CMB provided by the Planck Collaboration (2018)\cite{Planck:2018vyg}. In our analysis, we use both the high-$\ell$ and low-$\ell$ likelihoods, as well as the lensing likelihood:

\textbf{High-$\ell$ TT/TE/EE}: \texttt{Planck\_high\_l\_TTTEEE} covering $30 < \ell < 2508$ for TT and $30 < \ell < 1996$ for TE and EE.
     
\textbf{Low-$\ell$ TT}: \texttt{Planck\_lowl\_TT} for temperature in the range $2 < \ell < 29$.
    
\textbf{Low-$\ell$ EE}: \texttt{Planck\_lowl\_EE} for polarization in the same multipole range.
     
\textbf{Lensing} \cite{Planck:2018lbu}: \texttt{Planck\_lensing}, providing measurements of the projected gravitational potential derived from CMB lensing, offering an independent constraint on the growth of structure.

\subsection*{Baryon Acoustic Oscillations (BAO)}
For BAO measurements, we use the data from the Dark Energy Spectroscopic Instrument (DESI) Year Two dataset.\cite{DESI:2025zgx} In our analysis, we employ the corresponding likelihood \texttt{bao.desi\_dr2.desi\_bao\_all} in Cobaya. This dataset provides measurements of the Hubble parameter,the transverse comoving distance and the angle-averaged distance.

\begin{table}[t!]
\centering
\caption{Cosmological parameters and their priors used in the analysis.}
\begin{tabular}{|c|c|}
\hline
\textbf{Parameter} & \textbf{Prior} \\
\hline
$\Omega_bh^2$ & $ U[0.005, 0.03]$ \\
$\Omega_ch^2$ & $ U[0.01, 0.2]$ \\
$100\theta_s$ & $ U[0.5, 2]$ \\
$ln(10^{10}A_s)$ & $ U[2, 4]$ \\
$n_s$ & $U[0.7, 1.2]$ \\
$\tau_{reio}$ & $ U[0.0, 0.1]$ \\

\hline
\end{tabular}
\label{tab:priors}
\end{table}

\section{Results}\label{sec:res}
\renewcommand{\arraystretch}{1.4}

\begin{table}[H]
\centering
\caption{\small 68\% parameter intervals for $\Lambda$CDM model from previous data sets.}
\resizebox{\textwidth}{!}{
\begin{tabular}{lcccccc}
\toprule
Parameter & Planck & Planck+DESI & Planck+DESI+lensing & Planck+DESI+lensing+Union3 & Planck+DESI+lensing+PantheonPlus & Planck+DESI+lensing+DES \\
\midrule
$\Omega_{b}h^2$ & $0.02234\pm0.00015$ & $0.02252\pm0.00013$ & $0.02253\pm0.00012$ & $0.022250\pm0.00013$ & $0.02250\pm0.00013$ & $0.02249\pm0.00013$ \\
$\Omega_ch^2$  & $0.1202\pm0.0013$ & $0.11753\pm0.00067$ & $0.11767\pm0.00063$ & $0.1178\pm0.00065$ & $0.11789\pm0.00065$ & $0.11810\pm0.00064$ \\
$100\theta_s$  & $1.04185\pm0.00030$ & $1.04209\pm0.00027$ & $1.04209^{+0.00029}_{-0.00026}$ & $1.04209\pm0.00028$ & $1.04208\pm0.00028$ & $1.04204\pm0.00028$\\
$n_s$ & $0.9643\pm0.0043$ & $0.9707\pm0.0033$ & $0.9704\pm0.0032$ & $0.9697\pm0.0035$ & $0.9697\pm0.0033$ & $0.9694\pm0.0033$ \\
$\ln(10^{10}A_s)$ & $3.044\pm0.016$ & $3.046\pm0.017$ & $3.052\pm0.015$ & $3.052\pm0.015$ & $3.052\pm0.015$ & $3.051\pm^{+0.014}_{-0.016}$ \\
$\tau_{reio}$ & $0.0540\pm0.0078$ & $0.0578^{+0.0071}_{-0.0083}$ & $0.0606\pm0.0071$ & $0.0599\pm0.0073$& $0.0601\pm0.0074$ & $0.0595^{+0.0070}_{-0.0078}$ \\
\midrule
$H_0\ \mathrm{(km\,s^{-1}\,Mpc^{-1})}$ & $67.28\pm0.6$ & $68.46\pm0.30$ & $68.42\pm0.29$ & $68.32\pm0.29$ & $68.31\pm0.30$ & $68.21\pm0.28$ \\
$\Omega_{m0}$ & $0.3162\pm0.0086$ & $0.3002\pm0.0038$ & $0.3009\pm0.0036$ & $0.3022\pm0.0037$ & $0.3021\pm0.0037$ & $0.3035\pm0.0036$ \\
$S_8$ & $0.833\pm0.016$ & $0.8049\pm0.0098$ & $0.8089\pm0.0086$ & $0.8110\pm0.0083 $& $0.8112\pm0.0084$ & $0.8132\pm0.0083$ \\
$\sigma_8$ & $0.8113\pm0.0074$ & $0.8046^{+0.0066}_{-0.0074}$ & $0.8077\pm0.0062$ & $0.8081\pm0.0061$ & $0.8082\pm0.0061$ & $0.8084\pm0.0062$ \\
\midrule
$\chi^2$ & $2808.649$ & $2825.2674$ & $2834.4092$ & $2862.9566$ & $4242.1026$ & $4484.5508$ \\
\bottomrule
\end{tabular}
\label{tab:LCDM}

}
\end{table}

\renewcommand{\arraystretch}{1.4}


\renewcommand{\arraystretch}{1.4}

\begin{table}[H]
\centering
\caption{
\small
68\% parameter intervals for $f^{(I)}$CDM model from previous data sets
}
\resizebox{\textwidth}{!}{
\begin{tabular}{lcccccc}
\toprule
Parameter & Planck & Planck+DESI & Planck+DESI+lensing & Planck+DESI+lensing+Union3 & Planck+DESI+lensing+PantheonPlus & Planck+DESI+lensing+DES \\
\midrule
$\Omega_{b}h^2$ & $0.02244\pm0.00015$ & $0.02230\pm0.00012$ & $0.02230\pm0.00012$ & $0.02226\pm0.00013$ & $0.02220\pm0.00013$ & $0.02219\pm0.00013$ \\
$\Omega_ch^2$  & $0.1192\pm0.0013$ & $0.12088\pm0.00063$ & $0.12049\pm0.00064$ & $0.12101\pm0.00064$ & $0.12135\pm0.00062$ & $0.12172\pm0.00062$ \\
$100\theta_s$  & $1.04193\pm0.00030$ & $1.04180\pm0.00027$ & $1.04180\pm0.00028$ & $1.04179\pm0.00028$ & $1.04176\pm0.00028$ & $1.04175\pm0.00027$ \\
$n_s$ & $0.9672\pm0.0043$ & $0.9632\pm0.0033$ & $0.9639\pm0.0033$ & $0.9628\pm0.0031$ & $0.9618\pm0.0033$ & $0.9609\pm0.0033$ \\
$\ln(10^{10}A_s)$ & $3.036\pm0.016$ & $3.035\pm0.015$ & $3.017^{+0.016}_{-0.013}$ & $3.015^{+0.016}_{-0.013}$ & $3.013^{+0.017}_{-0.013}$ & $3.010^{+0.019}_{-0.013}$ \\
$\tau_{reio}$ & $0.0513\pm0.0077$ & $0.0489\pm0.0073$ & $0.0417^{+0.0080}_{-0.0064}$ & $0.0401^{+0.0081}_{-0.0061}$ & $0.0388^{+0.0088}_{-0.0065}$ & $0.0371^{+0.0100}_{-0.0062}$ \\
\midrule
$H_0\ \mathrm{(km\,s^{-1}\,Mpc^{-1})}$ & $73.43\pm0.69$ & $72.55\pm0.32$ & $72.72\pm0.32$ & $72.46\pm0.32$ & $72.27\pm0.32$ & $72.09\pm0.31$ \\
$\Omega_{m0}$ & $0.2641\pm0.0071$ & $0.2732\pm0.0034$ & $0.2713\pm0.0034$ & $0.2742\pm0.0034$ & $0.2762\pm0.0034$ & $0.2782\pm0.0034$ \\
$S_8$ & $0.823\pm0.016$ & $0.8414\pm0.0093$ & $0.8300\pm0.0081$ & $0.8350\pm0.0086$ & $0.8380\pm0.0085$ & $0.8408\pm0.0089$ \\
$\sigma_8$ & $0.8773\pm0.0080$ & $0.8817\pm0.0070$ & $0.8729^{+0.0067}_{-0.0059}$ & $0.8735^{+0.0071}_{-0.0058}$ & $0.8734^{+0.0074}_{-0.0061}$ & $0.8732^{+0.0082}_{-0.0058}$ \\
\midrule
$\chi^2$ & $2808.41$ & $2842.7966$ & $2855.0972$ & $2913.215400$ & $4309.051$ & $4574.3656$ \\
$\Delta\chi^2$ & $-0.239$ & $17.5292$ & $20.6880$ & $50.2588$ & $66.9484$ & $89.8148$ \\
\bottomrule
\end{tabular}
\label{tab:f2}

}
\end{table}


\renewcommand{\arraystretch}{1.5}
\begin{table}[H]
\centering
\caption{\small 68\% parameter intervals for $f^{(II)}$-CDM model from previous data set}
\label{tab:f3}
\begin{adjustbox}{max width=\textwidth}
\begin{tabular}{lcccccc}
\toprule
Parameter & Planck & Planck+DESI & Planck+DESI+lensing & Planck+DESI+lensing+Union3 & Planck+DESI+lensing+PantheonPlus & Planck+DESI+lensing+DES \\
\midrule
$\Omega_{b}h^2$ & $0.02229\pm0.00015$ & $0.01989\pm0.00010$ & $0.02000\pm0.00011$ & $0.019978^{+0.00009}_{-0.00010}$ & $0.01997\pm0.00011$ & $0.01996\pm0.00010$ \\
$\Omega_ch^2$ & $0.1191\pm0.0014$ & $0.15770\pm0.00083$ & $0.15407\pm0.00076$ & $0.15432\pm0.00081$ & $0.15448\pm0.00079$ & $0.15459\pm0.00074$ \\
$100\theta_s$ & $1.04191\pm0.00030$ & $1.03898\pm0.00028$ & $1.03892\pm0.00029$ & $1.03888\pm0.00028$ & $1.03888\pm0.00028$ & $1.03887\pm0.00030$ \\
$n_s$ & $0.97179\pm0.0046$ & $0.8997\pm0.0033$ & $0.9045\pm0.0032$ & $0.9034\pm0.0032$ & $0.9041\pm0.0033$ & $0.9035\pm0.0031$ \\
$\ln(10^{10}A_s)$ & $3.013^{+0.043}_{-0.053}$ & $3.0232^{+0.0064}_{-0.0071}$ & $2.9973\pm0.0056$ & $2.9973\pm0.0056$ & $2.9981\pm0.0057$ & $2.9975\pm0.0056$ \\
$\tau_{reio}$ & $0.040^{+0.022}_{-0.031}$ & $0.0046^{+0.0011}_{-0.0028}$ & $0.00275^{+0.00026}_{-0.0011}$ & $0.00276^{+0.00028}_{-0.0011}$ & $0.00277^{+0.00023}_{-0.0011}$ & $0.00265^{+0.00022}_{-0.0010}$ \\
\midrule
$H_0\ \mathrm{(km\,s^{-1}\,Mpc^{-1})}$ & $95.76\pm0.56$ & $82.44\pm0.24$ & $83.33\pm0.24$ & $83.24\pm0.25$ & $83.19\pm0.25$ & $83.16\pm0.25$ \\
$\Omega_{m0}$ & $0.1550\pm0.0032$ & $0.2622\pm0.0025$ & $0.2515\pm0.0023$ & $0.2525\pm0.0023$ & $0.2529\pm0.0023$ & $0.2532\pm0.0023$ \\
$z_{tr}$ & $1.487\pm0.027$ & $0.9731\pm0.0089$ & $1.0136\pm0.0088$ & $1.007\pm0.0079$ & $1.016\pm0.012$ & $1.007\pm0.0079$ \\
$S_8$ & $0.729\pm0.017$ & $1.131\pm0.010$ & $1.0801\pm0.0088$ & $1.0831\pm0.0093$ & $1.0852\pm0.0090$ & $1.0858\pm0.0087$ \\
$\sigma_8$ & $1.014^{+0.023}_{-0.016}$ & $1.2095\pm0.0061$ & $1.1795\pm0.0051$ & $1.1806\pm0.0053$ & $1.1818\pm0.0051$ & $1.1816^{+0.0044}_{-0.0050}$ \\
\midrule
$\chi^2$ & $2934.745200$ & $4244.9736$ & $4389.9164$ & $4481.664$ & $5888.434$ & $6174.497800$ \\
$\Delta\chi^2$ & $126.0962$ & $1419.7062$ & $1555.5072$ & $1618.7074$ & $1646.3314$ & $1689.9470$ \\
\bottomrule
\end{tabular}
\end{adjustbox}
\end{table}

\begin{figure}
    \centering
    \includegraphics[width=0.8\linewidth]{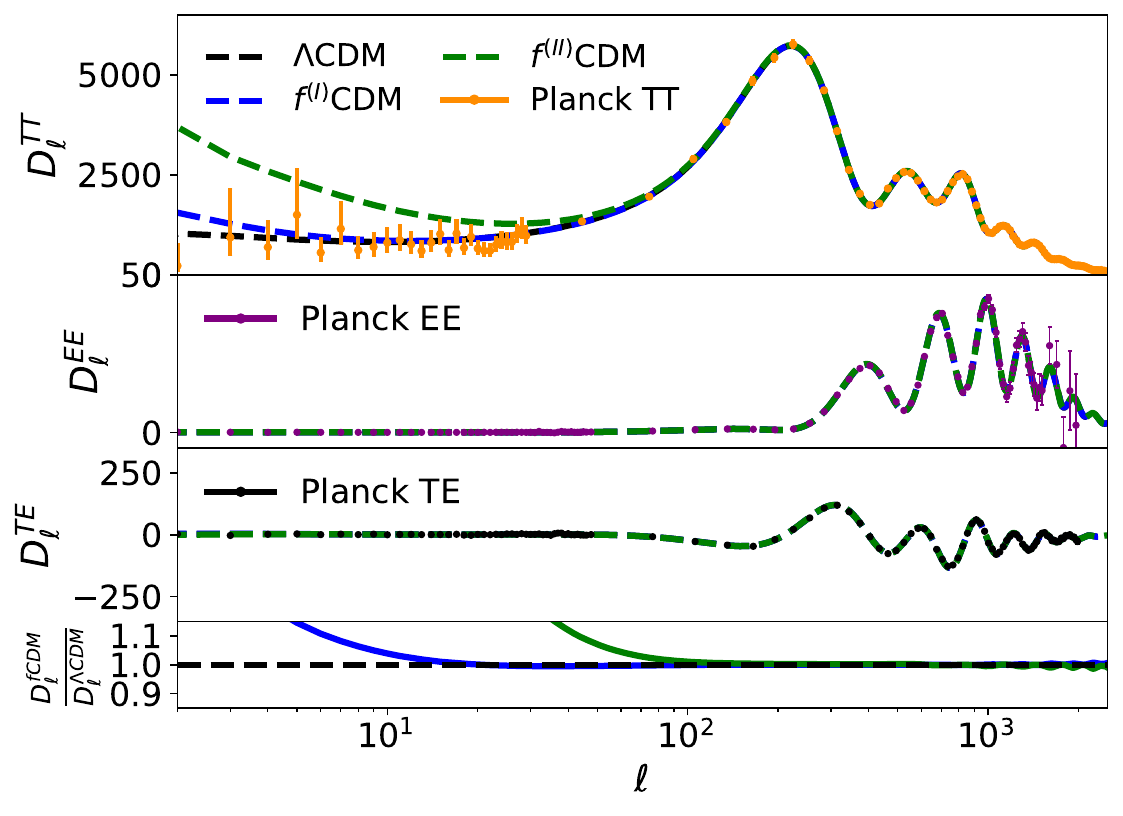}
    \caption{The CMB temperature, polarization anisotropies, and their cross-correlation ($TT$, $EE$, and $TE$) are presented for the $\Lambda$CDM (black dashed curve), $f^{(I)}$CDM (blue dashed curve), and $f^{(II)}$CDM (green dashed curve) models, plotted against the Planck data sets (circular points). The lower panel displays the ratio $C_{\ell}^{TT} / (C_{\ell}^{TT})_{\Lambda\text{CDM}}$ for the $f^{(I)}$ and $f^{(II)}$ models, using the best-fit parameters constrained by Planck data. It is observed that all three models exhibit nearly identical behavior across most multipoles, except at small $\ell$ (large scales). This discrepancy at low multipoles arises from the differences in the Integrated Sachs-Wolfe (ISW) effect among the three models, which affects large-scale modes entering the horizon during the dark energy dominated era.}
    \label{fig:plankdata}
\end{figure}
\subsection{$f^{(I)}\text{CDM}$ vs $\Lambda \text{CDM}$}
\begin{figure}[htbp]
    \centering
    \includegraphics[width=\textwidth]{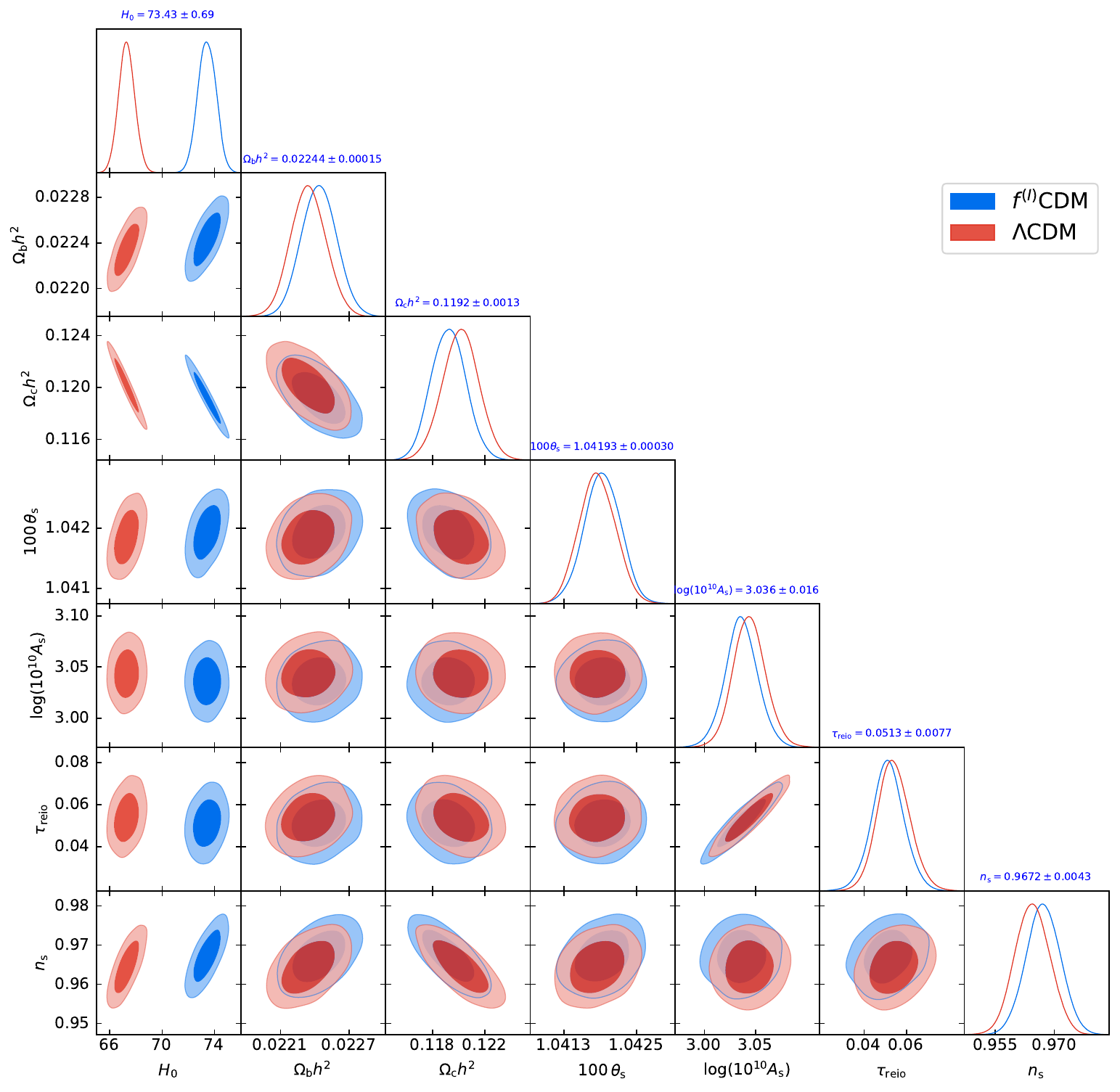}
    \caption{The 68\% (dark color) and 95\% (faint color) confidence level (CL) contours of the cosmological parameters for the $f^{(I)}$CDM VS $\Lambda$CDM models using Planck data set. We also include the one-dimensional posterior distributions for the parameters.}
    \label{fig:f2_planck_lambda}
\end{figure}

Tables \ref{tab:LCDM} and \ref{tab:f2} represents the best-fit values and $68\%$ confidence intervals for the primary cosmological parameters listed in Table~\ref{tab:priors}, as well as the four derived parameters $H_0$, $\sigma_8$, $\Omega_{m0}$ and $S_8$ for the $f^{(I)}\text{CDM}$ and $\Lambda\text{CDM}$ models. Since this $f^{(I)}\text{CDM}$ model mimics the $\Lambda\text{CDM}$ behavior in the early Universe, for the Planck dataset the primary parameters shift by less than $1\sigma$ from their best-fit values, as shown in Fig.~\ref{fig:f2_planck_lambda}. 
This behavior is further illustrated in Fig.~\ref{fig:plankdata}, where the $f^{(I)}$ predictions for the CMB anisotropy power spectra ($TT$, $EE$, and $TE$) are nearly identical to those of $\Lambda\text{CDM}$, except at low multipoles ($\ell$) in the $C^{TT}_\ell$ spectrum. In this region, the $f^{(I)}$ curve exhibits a slightly larger amplitude than $\Lambda\text{CDM}$. This discrepancy is attributed to the modified Late Integrated Sachs-Wolfe (LISW) effect, which influences large-scale perturbations at late times depending on the evolution of the gravitational potentials $\Phi$ and $\Psi$. This modified dynamics can be traced to the gravitational slip, as defined in Eq.~\eqref{eq:slib}, which significantly affects larger scales (low $\ell$).

\begin{figure}[t!]
    \centering
    \begin{subfigure}{0.48\textwidth} 
        \centering
        \includegraphics[width=\linewidth]{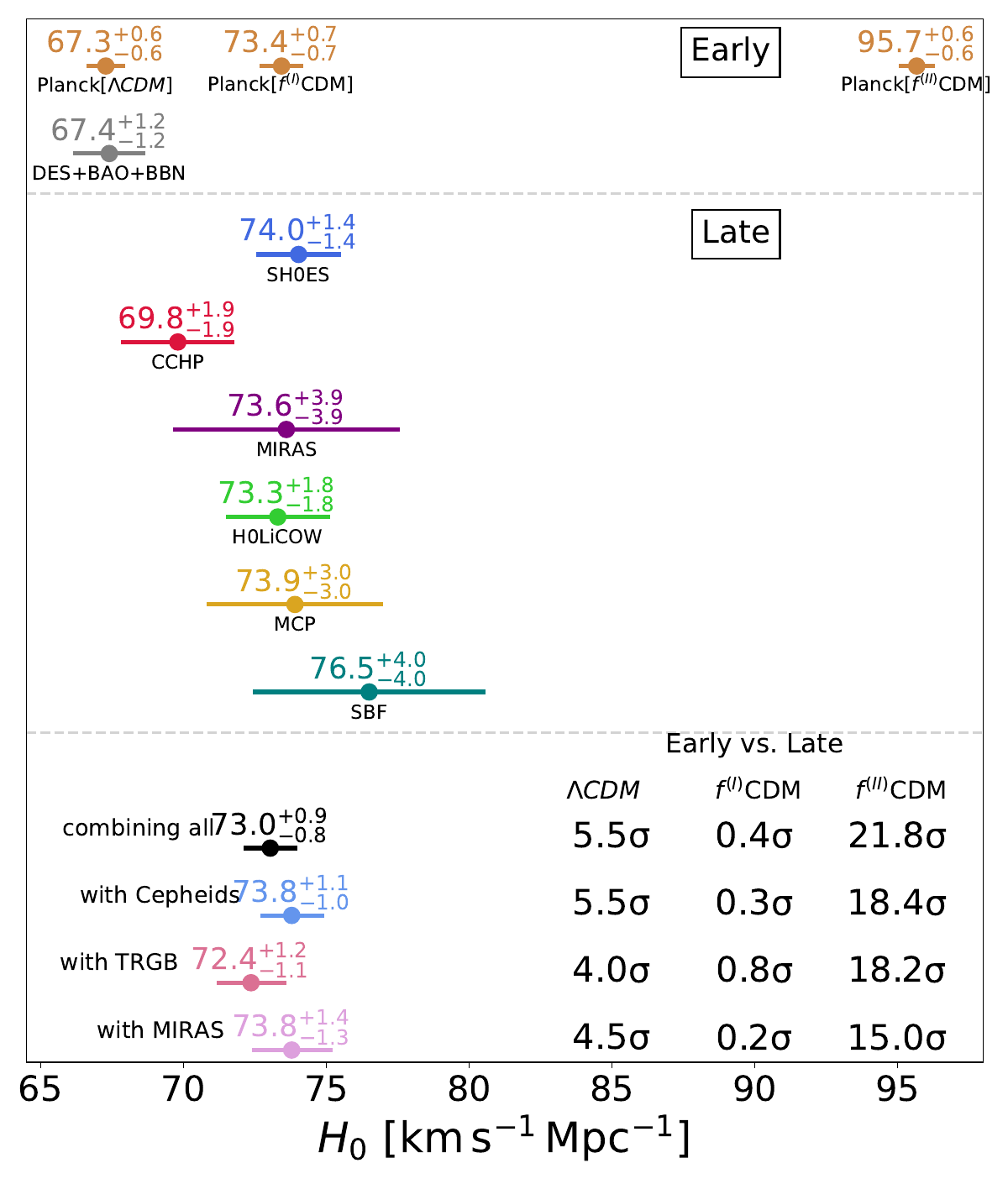}
        \caption{H0 tension.\protect\footnotemark}\label{fig:H0tns}
    \end{subfigure}
    \hfill 
    \begin{subfigure}{0.48\textwidth}
        \centering
        \includegraphics[width=\linewidth]{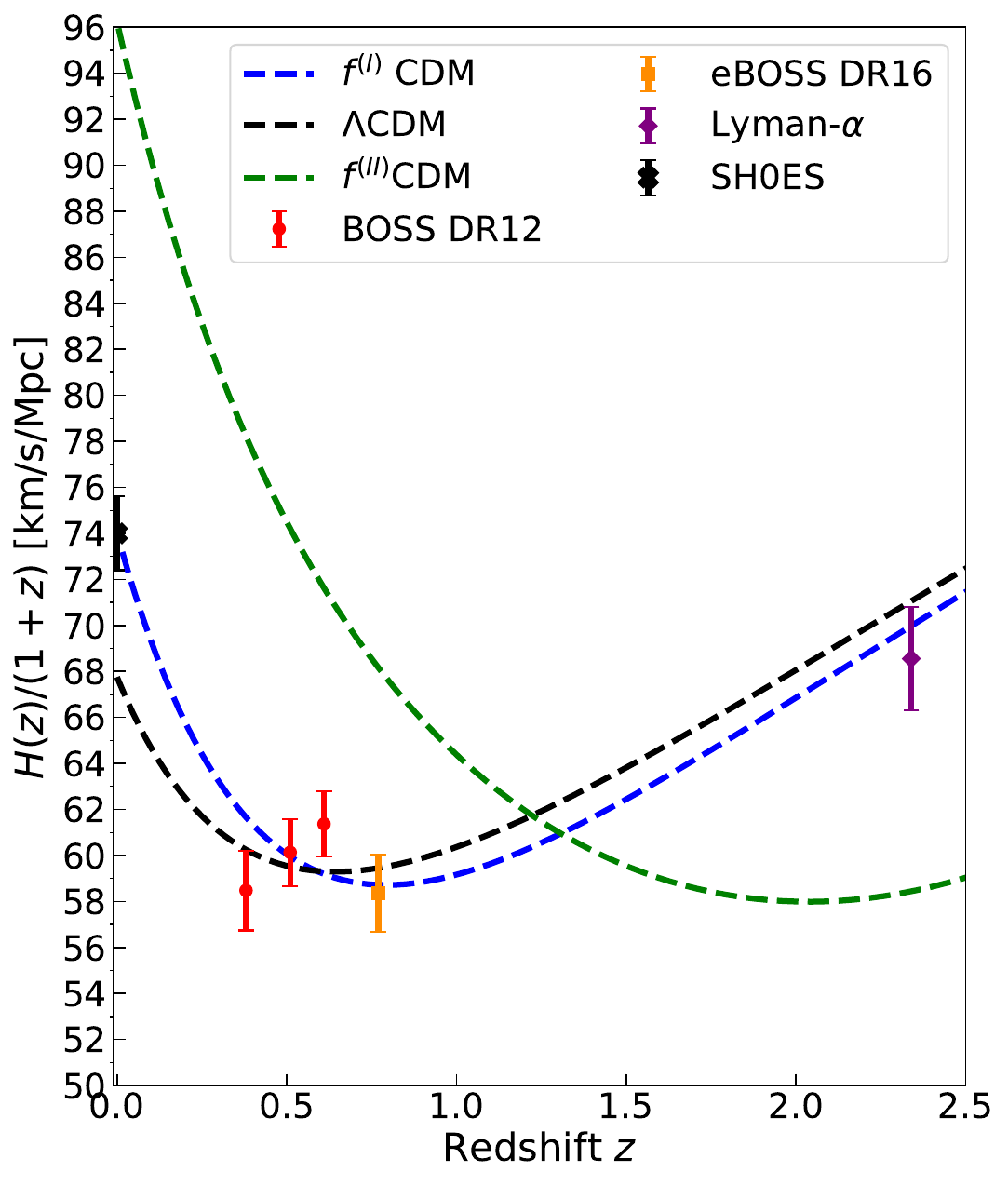}
        \caption{Evolution of the Hubble parameter.} \label{fig:hbbl}
    \end{subfigure}
    \caption{Using the best-fit cosmological parameters derived from the \textit{Planck} data set. (\subref{fig:H0tns}) Comparison of the $H_0$ constraints from \textit{Planck} for the $\Lambda$CDM, $f^{(I)}$CDM, and $f^{(II)}$CDM models with various local measurements. The $f^{(I)}$ model brings the \textit{Planck} prediction into agreement with all local $H_0$ determinations. (\subref{fig:hbbl}) Evolution of $H(z)/(1+z)$ for the $\Lambda$CDM (black), $f^{(I)}$CDM (blue), and $f^{(II)}$CDM (green) models over $0\le z\le2.5$, compared with BAO measurements from BOSS DR12 \cite{BOSS:2016wmc}, eBOSS DR16 \cite{eBOSS:2020xwt}, the eBOSS DR14 Lyman-$\alpha$ forest data \cite{eBOSS:2019ytm}, and the SH0ES local $H_0$ measurement.
} 
    \label{fig:hubble}
\end{figure}

Regarding the derived parameters, the model $f^{(I)}\text{CDM}$  predicts a significantly larger Hubble constant, $H_0 = 73.43 \pm 0.69 \text{ km s}^{-1} \text{ Mpc}^{-1}$, compared to the $\Lambda\text{CDM}$ value of $H_0 = 67.28 \pm 0.60 \text{ km s}^{-1} \text{ Mpc}^{-1}$. This result effectively resolves the tension with local measurements of $H_0$ \cite{Riess:2016jrr}, as shown in Fig.~\ref{fig:H0tns}, where the discrepancy is reduced to less than $0.8\sigma$ for all local datasets. This elevated $H_0$ value for $f^{(I)}$ when using Planck data was expected, given its phantom-like behavior at late times (See Fig.\ref{fig:DEOS}), which is a known mechanism for increasing the Hubble constant \cite{Ludwick:2017tox}.

\begin{figure}
    \centering
    \includegraphics[width=0.6\linewidth]{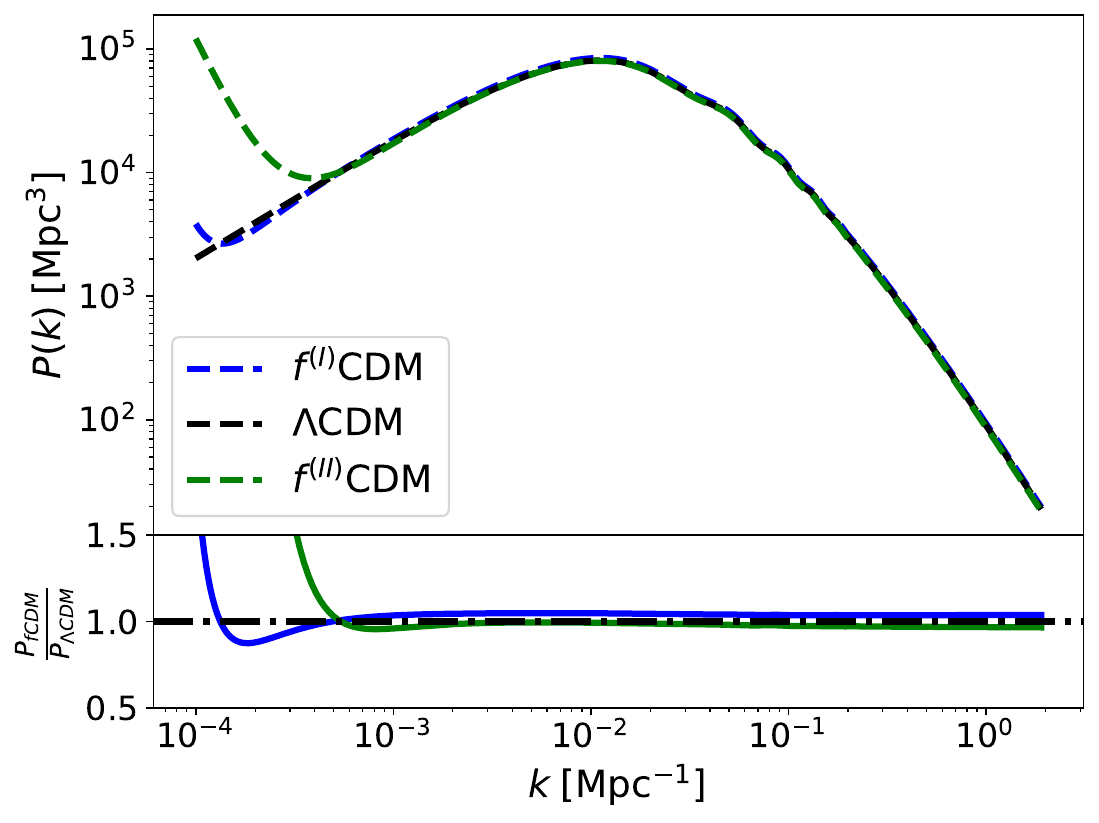}
    \caption{The matter power spectrum $P(k)$ is presented for the $\Lambda \text{CDM}$ model (black dashed curve), alongside the $f^{(I)}\text{CDM}$ (blue dashed curve) and $f^{(II)}\text{CDM}$ (green dashed curve) models. The lower panel illustrates the ratio $P(k)/P_{\Lambda}(k)$ for both $f^{(I)}, f^{(II)}\text{models}$  using bestfit values of cosmological parameters from Plank data set. It is observed that the $f^{(I)}\text{model}$  exhibits a larger $P(k)$ across almost all modes $k$ that entered the horizon during the radiation and matter-dominated eras. In contrast, the $f^{(II)}\text{model}$  remains almost identical to $\Lambda \text{CDM}$ for these modes. For large-scale modes that enter the horizon at late times (during the dark energy dominated era), the power spectrum for $f^{(II)}, f^{(I)}$ initially becomes smaller than that of $\Lambda \text{CDM}$, before transitioning to significantly larger values at the smallest $k$ modes (largest scales).}
    \label{fig:mttrpwr}
\end{figure}

Furthermore, the $f^{(I)}\text{model}$  yields a best-fit value of $\sigma_8 = 0.8775 \pm 0.0080$, with \text{Planck} data, which is slightly higher than the $\Lambda\text{CDM}$ value ($\sigma_8 = 0.8113 \pm 0.0074$). This improvement can be attributed to two main factors: first, the higher value of $H_0$ in the $f^{(I)}\text{model}$ ; and second, the increased growth of structures as shown in Fig.~\ref{fig:mttrpwr}, where the matter power spectrum $P(k)$ for $f^{(I)}$ is amplified due to a more efficient growth rate. This increased growth is a direct consequence of the stronger gravitational interaction in the $f^{(I)}\text{model}$, characterized by an effective gravitational constant $G_{\text{eff}}=GQ$ that exceeds the standard Newtonian value. For further details on this mechanism, see \cite{Bedair:2026M4}.

When the DESI data are combined with the Planck dataset, the best-fit of most primary parameters for the $f^{(I)}\text{CDM}$ model remains consistent with $\Lambda\text{CDM}$, with discrepancies staying below the $2\sigma$ level. However, the Hubble constant decreases slightly to $H_0 = 72.55 \pm 0.32 \text{ km s}^{-1} \text{ Mpc}^{-1}$. Despite this decrease, the model continues to significantly reduce the tension with local $H_0$ observations compared to the standard $\Lambda\text{CDM}$ framework. 

An interesting feature emerges upon the inclusion of DESI data in the matter-density parameter, $\omega_{m0}$. A notable shift is observed relative to the Planck-only constraint. Interestingly, a similar shift is also present in the $\Lambda\mathrm{CDM}$ model, although in the opposite direction: $\omega_{m0}$ decreases in $\Lambda\mathrm{CDM}$, whereas it shifts toward larger values in the $f^{(I)}\mathrm{CDM}$ model. This difference is accompanied by an opposite shift in the scalar spectral index $n_s$. Such behavior can be understood in terms of the intrinsic $\omega_{m0}$--$n_s$ degeneracy in the Planck likelihood \cite{Efstathiou:1998xx,Ferreira:2025lrd}. Consequently, a DESI-induced shift in $\omega_{m0}$ propagates into the preferred value of $n_s$ along this degeneracy direction. For a detailed discussion of the this behavior, see Ref.~\cite{Bedair:2026M4}.

\begin{figure}[htbp]
    \centering
    \includegraphics[width=\textwidth]{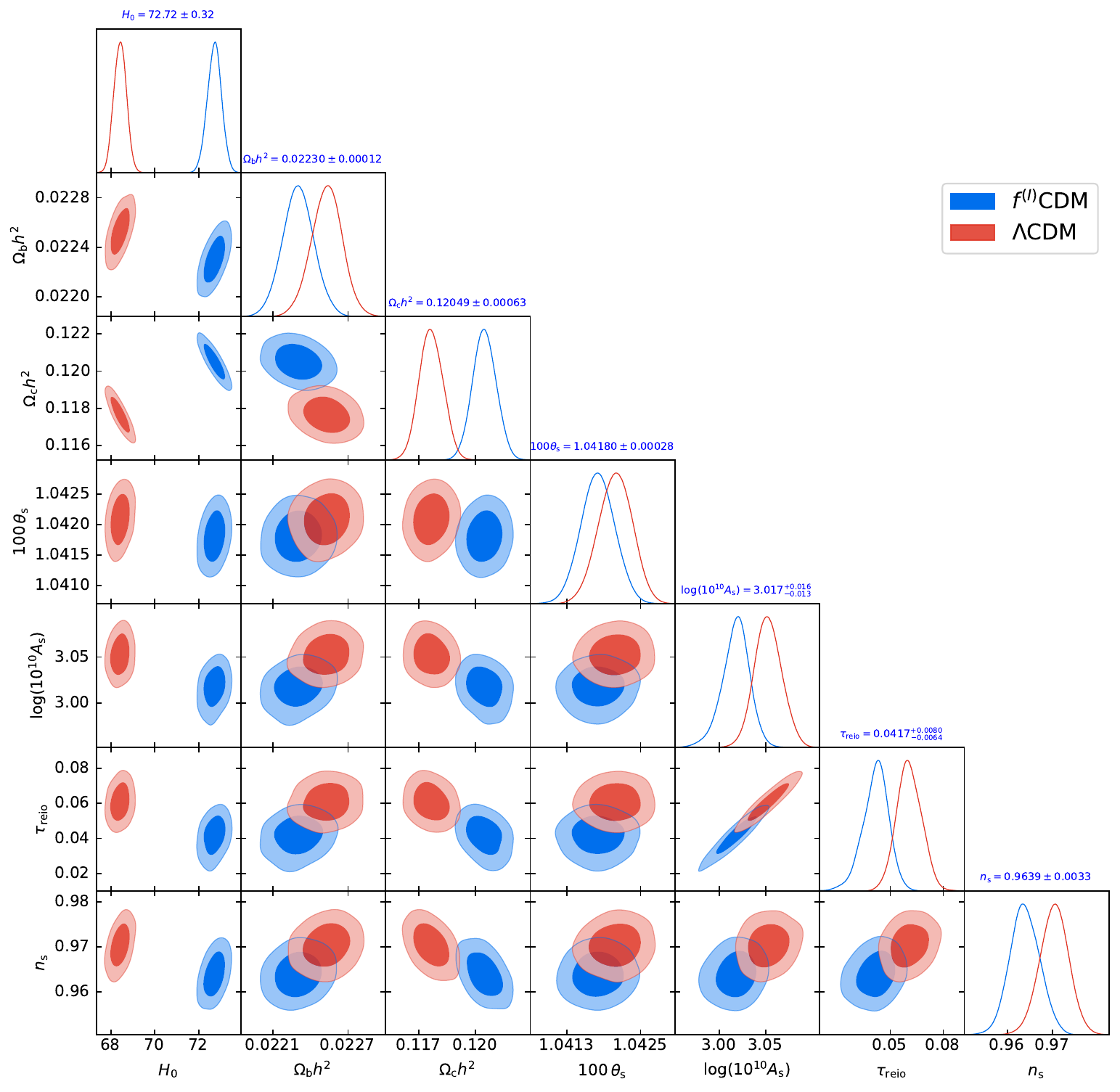}
    \caption{The 68\% (dark color) and 95\% (faint color) confidence level (CL) contours of the cosmological parameters for the $f^{(I)}$CDM VS $\Lambda$CDM models using \text{Planck+Desi+lensing} data set. We also include the one-dimensional posterior distributions for the parameters.}
    \label{fig:f2_len_lambda}
\end{figure}

A more interesting behavior is observed in the reionization optical depth,$\tau_{\text{reio}}$, and the primordial amplitude, $A_s$. As illustrated in Fig.~\ref{fig:f2_len_lambda}, the inclusion of CMB lensing data induces an opposite shift in these parameters in the $f^{(I)}\text{CDM}$ model compared with $\Lambda$CDM. In particular, the $f^{(I)}\text{CDM}$ model prefers a lower value of $A_s$. This behavior can be understood from the enhanced growth of structure predicted by the model (see Fig.~\ref{fig:mttrpwr}). Since the enhanced growth increases the amplitude of the matter fluctuations and, consequently, the lensing signal, consistency with the CMB lensing measurements can be achieved through a lower primordial amplitude $A_s$. The resulting shift in $A_s$ is also reflected in the inferred value of $\tau_{\text{reio}}$. The high-$\ell$ CMB anisotropies primarily constrain the combination $A_s e^{-2\tau_{\text{reio}}}$, which approximately determines the amplitude of the observed acoustic peaks. Therefore, a decrease in $A_s$ is accompanied by a corresponding decrease in $\tau_{\text{reio}}$, preserving the combination constrained by the high-$\ell$ CMB data.

\begin{figure}[htbp]
    \centering
    \includegraphics[width=\textwidth]{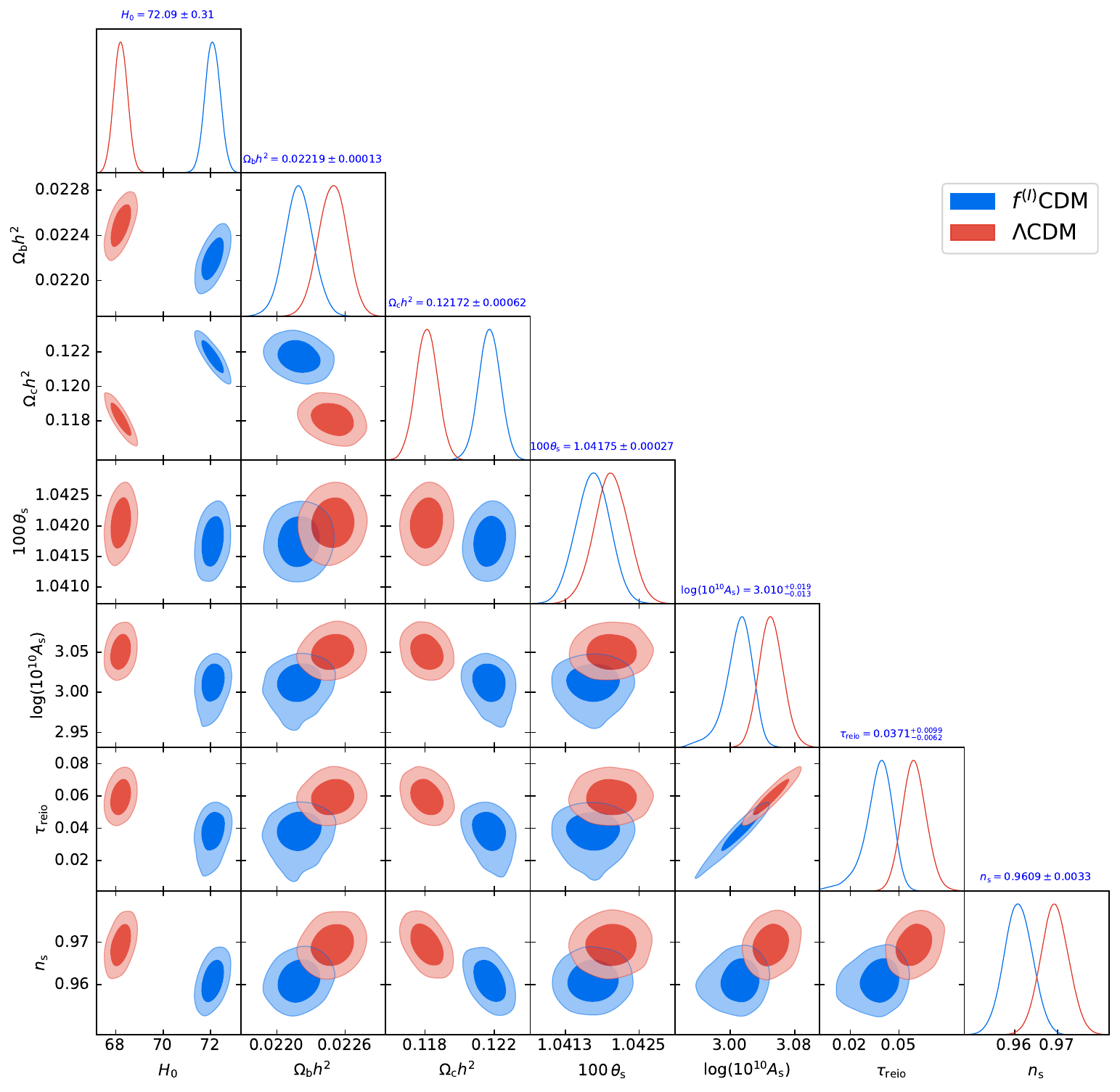}
    \caption{The 68\% (dark color) and 95\% (faint color) confidence level (CL) contours of the cosmological parameters for the $f^{(I)}$CDM VS $\Lambda$CDM models using \text{Planck+Desi+lensing+DES} data set. We also include the one-dimensional posterior distributions for the parameters.}
    \label{fig:f2_DES_lambda}
\end{figure}

In contrast, the addition of Type Ia Supernova (SNIa) data does not significantly alter the best-fit parameters, as shown in Fig.~\ref{fig:f2_DES_lambda}. The results of the combined datasets remain largely compatible with the $\text{Plank+DESI+LEN}$ data set, and still accommodate the modified gravity features of the $f^{(I)}\text{model}$.

Since the $f^{(I)}$CDM model has the same number of independent cosmological parameters as $\Lambda$CDM, their statistical comparison is straightforward and can be based directly on the difference in the best-fit $\chi^2$. As shown in Table~\ref{tab:f2}, the $f^{(I)}$CDM model is mildly preferred over $\Lambda$CDM when constrained by the \textit{Planck}-only dataset. However, this preference is reversed once background observations, including DESI BAO and Type Ia supernovae, are incorporated. In particular, the addition of the background data leads to a substantial increase in the relative $\chi^2$ of the $f^{(I)}$CDM model, resulting in a significantly stronger preference for $\Lambda$CDM.

\footnotetext{Figure~\ref{fig:hubble}(\subref{fig:H0tns}) is generated after modifying the code introduced by Vivien Bonvin and Martin Millon, available at \href{https://github.com/vbonvin/H0_tension} {\faGithub} \url{https://github.com/vbonvin/H0_tension}, in order to implement the models discussed in this work.}

\clearpage

\subsection{$f^{(II)}$CDM vs $\Lambda$CDM}

Table~\ref{tab:f3} summarizes the best-fit values and the 68\% confidence intervals for the primary cosmological parameters together with the derived quantities $H_0$, $\Omega_{m0}$, $\sigma_8$, $S_8$, and $z_{tr}$. Using the \textit{Planck} dataset alone, the $f^{(II)}$CDM model predicts a remarkably high Hubble constant,
\[
H_0=95.76\pm0.56~{\rm km\,s^{-1}\,Mpc^{-1}},
\]
which corresponds to a disagreement exceeding $\sim18\sigma$ with the local SH0ES determination (see Fig.~\ref{fig:hubble}\subref{fig:H0tns}), substantially worsening the $H_0$ tension relative to $\Lambda$CDM.

\begin{figure}[htbp]
    \centering
    \includegraphics[width=\textwidth]{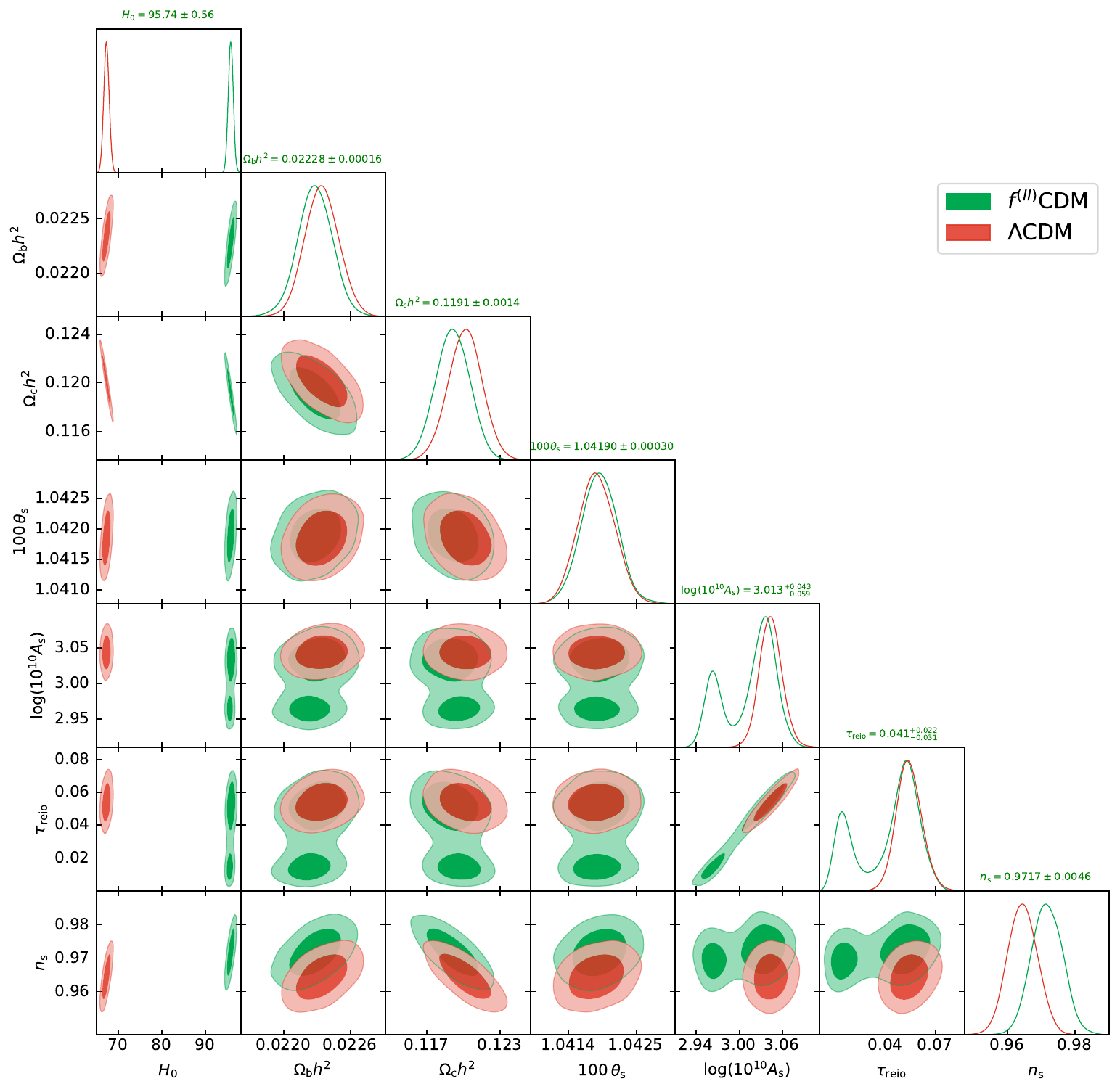}
    \caption{The 68\% (dark color) and 95\% (faint color) confidence level (CL) contours of the cosmological parameters for the $f^{(II)}$CDM VS $\Lambda $CDM models using \text{Planck} data set We also include the one-dimensional posterior distributions for the parameters.}
    \label{fig:f3_plank_lambda}
\end{figure}

A distinctive feature of the model is the bimodal posterior distribution of the primordial parameters $\tau_{\rm reio}$ and $A_s$, shown in Fig.~\ref{fig:f3_plank_lambda}. This behavior originates from the internal competition between different components of the \textit{Planck} likelihood. At high multipoles, the CMB spectra constrain mainly the combination $A_s e^{-2\tau_{\rm reio}}$, leaving the two parameters highly degenerate. This degeneracy is  broken by the low-$\ell$ temperature and polarization likelihoods, which will constrain $A_s$ in different ways

In the $f^{(II)}$ model, the late-time evolution of the gravitational potentials enhances the Late Integrated Sachs--Wolfe (LISW) contribution relative to $\Lambda$CDM (see Fig.~\ref{fig:LISW}). The additional large-scale TT power drives the low-$\ell$ TT likelihood toward smaller values of $A_s$ in order to compensate for the enhanced LISW signal. Through the high-$\ell$ degeneracy between $A_s$ and $\tau_{\rm reio}$, this shift also favors smaller values of $\tau_{\rm reio}$, producing the secondary mode of the posterior distribution(See Fig \ref{fig:Planck+lowT}). Conversely, the low-$\ell$ EE polarization data strongly prefer the standard reionization solution with larger $\tau_{\rm reio}$(See Fig \ref{fig:Planck+lowE}). The competition between these two inconsistent constraints naturally generates the observed bimodal posterior (See Fig \ref{fig:Planck+lowT+lowE}).

The remaining primary cosmological parameters remain compatible with their $\Lambda$CDM values within approximately $1\sigma$. This agreement is expected because the $f^{(II)}$ model closely follows the $\Lambda$CDM expansion history prior to the late Universe, where the CMB acoustic peaks are formed.

\begin{figure}[htbp]
    \centering
    \includegraphics[width=0.8\linewidth]{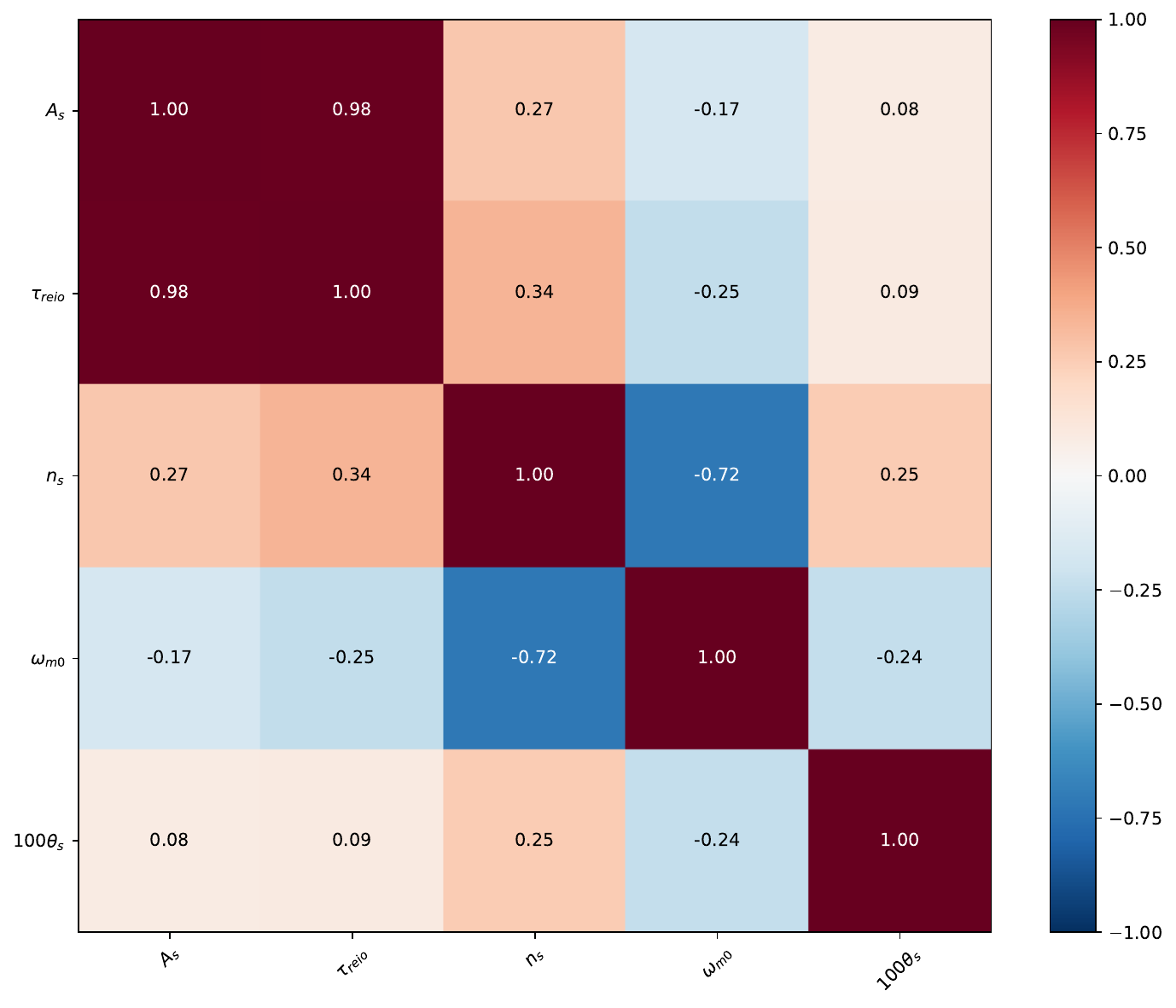}
    \caption{ The correlation matrix for the baseline cosmological parameters of the $f^{(II)}\text{model}$, derived from Planck 2018 measurements. The color scale indicates the strength of the statistical degeneracy between parameters, ranging from dark blue (representing a strong negative correlation, $-1$) to dark red (representing a strong positive correlation, $+1$). This matrix highlights the primary degeneracy directions, such as the well-known positive correlation between $\tau$ and $A_s$, and the negative correlation between $\omega_m$ and $n_s$, which characterize the parameter space of the modified gravity model. }
    \label{fig:crrmtrx}
\end{figure}

\begin{figure}[htbp]
    \centering
    \includegraphics[width=\textwidth]{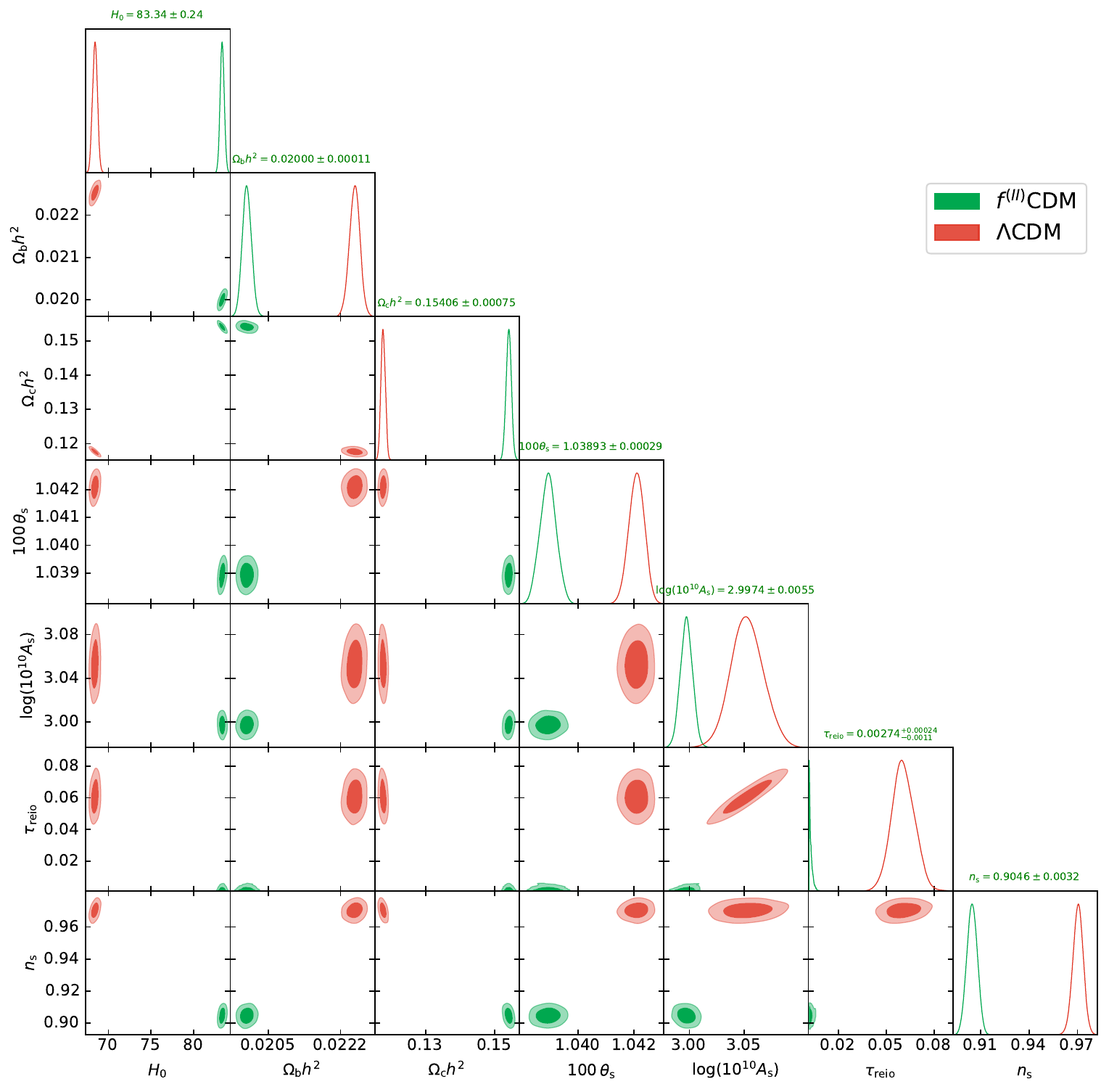}
    \caption{The 68\% (dark color) and 95\% (faint color) confidence level (CL) contours of the cosmological parameters for the $f^{(II)}$CDM VS $\Lambda $CDM models using \text{Planck+Desi+lensing} data set We also include the one-dimensional posterior distributions for the parameters.}
    \label{fig:f3_len_lambda}
\end{figure}

When DESI measurements are combined with the \textit{Planck} likelihood, the situation changes significantly (Fig.~\ref{fig:f3_len_lambda}). The additional BAO information strongly constrains the matter density, shifting the preferred value of $\omega_m$ to larger values. Owing to the intrinsic parameter degeneracies of the CMB likelihood (See Fig \ref{fig:crrmtrx}), this shift propagates to several primordial parameters, producing significant deviations from the corresponding $\Lambda$CDM constraints. In particular, the combined analysis yields
\[
100\theta_s=1.03898\pm0.00028,\qquad
\omega_m=0.17821\pm0.00082,\qquad
n_s=0.8997\pm0.0033,\qquad
\]
corresponding to approximately $7\sigma$, $23\sigma$, and $12\sigma$ departures from the $\Lambda$CDM values, respectively.

This behavior can be understood from the strong parameter correlations present in the \textit{Planck} likelihood (Fig.~\ref{fig:crrmtrx}). In particular, the correlation between $100\theta_s$ and $\omega_m$ approximately follows
\[
\delta(100\theta_s)\simeq-0.05538\,\delta\omega_m.
\]
Using the observed shift
$\delta\omega_m\simeq0.036$,
this relation predicts
$\delta(100\theta_s)\simeq-0.002$,
in good agreement with the numerical MCMC results shown in Table~\ref{tab:f3}. The remaining  discrepancy is expected because the relation is obtained from a linear approximation, whereas the full likelihood is mildly nonlinear.

Although the theoretical construction of the model is well motivated, its phenomenology leads to non-physical solutions in certain regions of parameter space. Interestingly, these regions are associated with a bimodal behavior of the inferred cosmological parameters, making the model an instructive case study.

\begin{figure}[htbp]
    \centering
    \includegraphics[width=\textwidth]{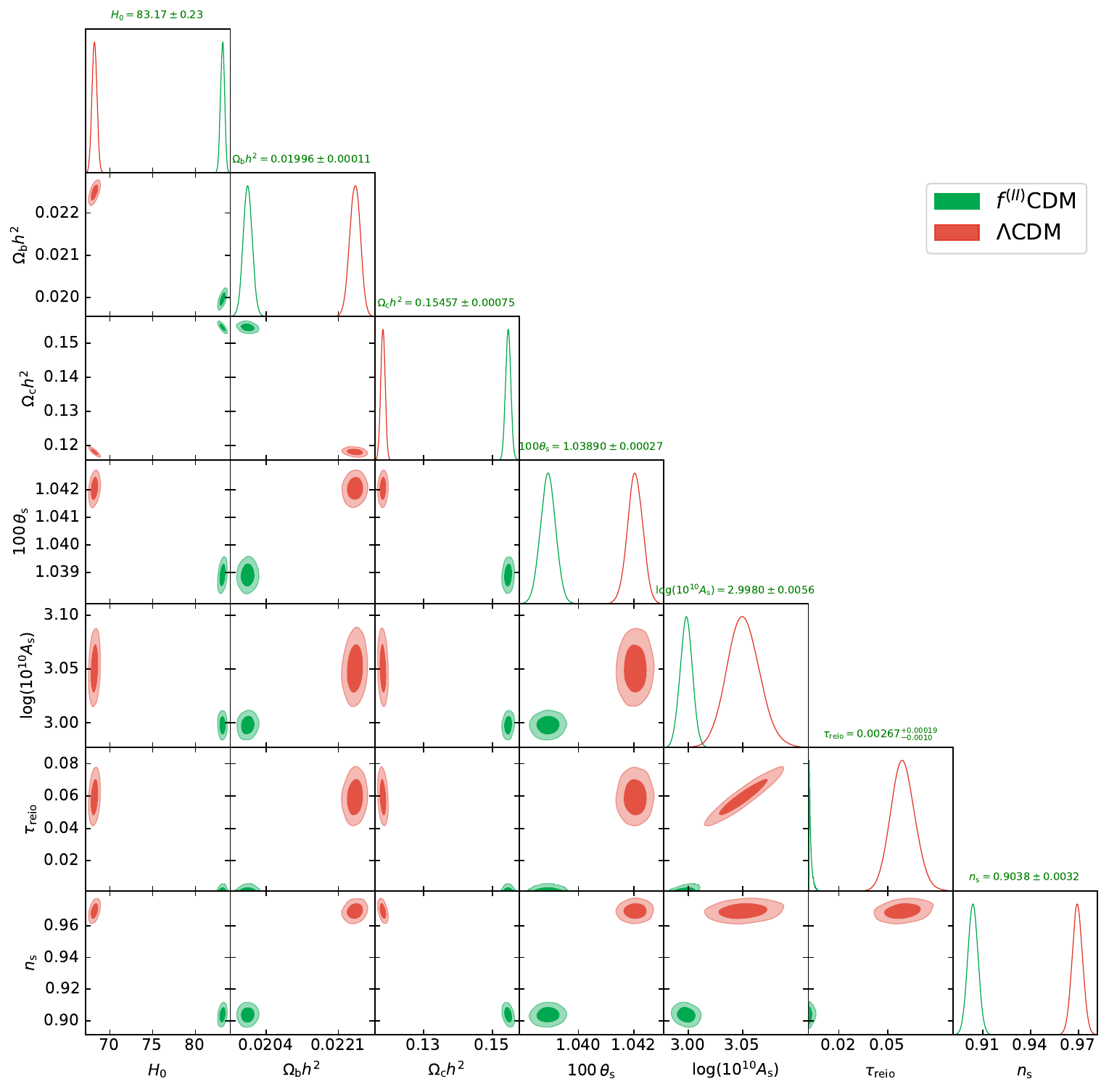}
    \caption{The 68\% (dark color) and 95\% (faint color) confidence level (CL) contours of the cosmological parameters for the $f^{(II)}$CDM VS $\Lambda $CDM models using \text{Planck+Desi+lensing+DES} data set We also include the one-dimensional posterior distributions for the parameters.}
    \label{fig:f3_DES_lambda}
\end{figure}

\clearpage

\subsection{The three model comparison}
\begin{figure}[htbp]
    \centering
    \includegraphics[width=\textwidth]{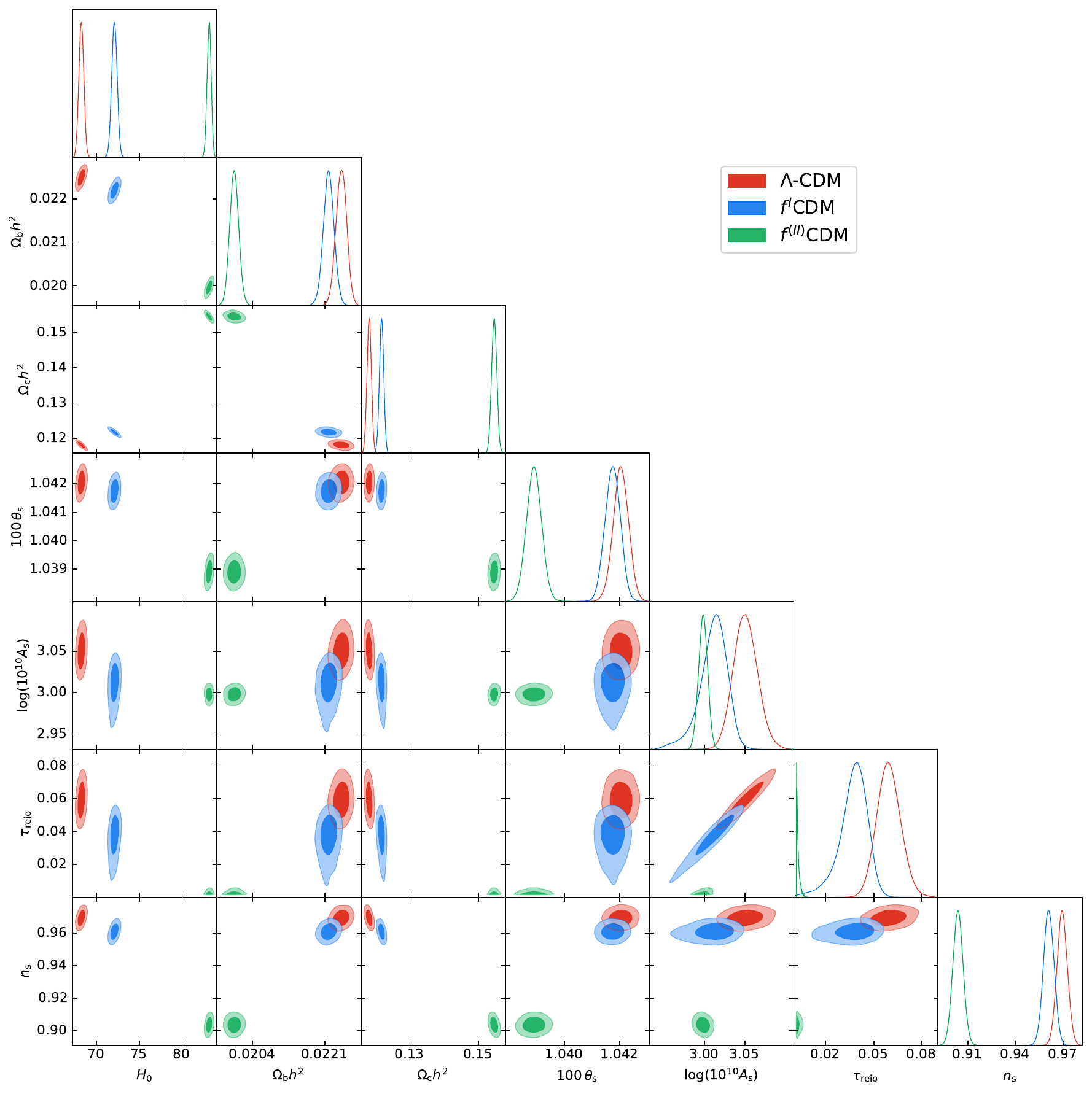}
    \caption{The 68\% (dark color) and 95\% (faint color) confidence level (CL) contours of the cosmological parameters for the $f^{(I)}$CDM VS $\Lambda $CDM VS $f^{(II)}$CDM  models using \text{Planck+Desi+lensing+DES} data set We also include the one-dimensional posterior distributions for the parameters.}
    \label{fig:f3_DES_lambda}
\end{figure}
\newpage

\section{Summary and Conclusion}\label{sec:con}

In this work, motivated by the new exponential infrared $f(T)$ teleparallel gravity model proposed in \cite{Hashim:2026M3}, we investigated the case corresponding to $\alpha=+1$. Owing to the two branches of the Lambert $W$ function, the model naturally gives rise to two distinct cosmological solutions, denoted by $f^{(I)}$ and $f^{(II)}$, each leading to fundamentally different cosmological evolutions.

The primary branch, $f^{(I)}$, exhibits a phantom-crossing dark-energy behavior. Consistent with the expected phenomenology of phantom models \cite{Ludwick:2017tox}, it predicts a relatively high Hubble constant, yielding $H_0=73.43\pm0.69~{\rm km\,s^{-1}\,Mpc^{-1}}$ using the \textit{Planck} dataset and $H_0=72.55\pm0.32~{\rm km\,s^{-1}\,Mpc^{-1}}$ after including DESI observations, substantially reducing the $H_0$ tension with local measurements. Although this branch is mildly preferred over $\Lambda$CDM using \textit{Planck} data alone, the inclusion of background observations reverses this preference, making $\Lambda$CDM statistically favored.

The secondary branch, $f^{(II)}$, follows a qualitatively different cosmological evolution. It predicts an effective dark-energy density that changes sign at late times, with a transition redshift of $z_{\rm tr}\sim1$. This transition occurs at a lower redshift not only compared to other negative or sign-changing dark-energy models \cite{Soriano:2025gxd,Gomez-Valent:2024ejh}, but also relative to the value ($z_{\rm tr}\approx1.6$) specifically proposed to resolve the Ly$\alpha$ BAO tension \cite{BOSS:2014hhw} via a high-redshift negative dark-energy density. Such scenarios have recently attracted considerable attention in light of the DESI observations \cite{DESI:2024aqx} and the early structure formation inferred from JWST data \cite{Menci:2024rbq}.

A particularly interesting result of the $f^{(II)}$ branch is the appearance of a bimodal posterior distribution for the primordial parameters $A_s$ and $\tau_{\rm reio}$ when constrained by \textit{Planck} data alone. We demonstrated that this bimodality originates from the internal tension between the low-$\ell$ TT and low-$\ell$ EE \textit{Planck} likelihoods. More generally, our analysis suggests that such behavior is expected in cosmological models that predict a sufficiently enhanced late Integrated Sachs--Wolfe contribution relative to $\Lambda$CDM.

Despite these interesting phenomenological features, the $f^{(II)}$ branch becomes cosmologically non-viable once DESI measurements are included. The background data strongly shift the preferred cosmological parameters through their interplay with the parameter degeneracies present in the \textit{Planck} likelihood, leading to severe inconsistencies with the corresponding $\Lambda$CDM constraints.

Overall, our analysis demonstrates that the two Lambert branches describe fundamentally different cosmological scenarios. The primary branch remains phenomenologically viable and significantly alleviates the $H_0$ tension while remaining compatible with current observations. In contrast, although the secondary branch provides an interesting realization of sign-changing dark energy and reveals a novel manifestation of the internal low-$\ell$ \textit{Planck} TT--EE tension, it is ultimately ruled out by current background cosmological observations.

\section{Acknowledgements}
This paper is supported by the Science, Technology \& Innovation Funding Authority (STDF), Egypt, under the Basic Sciences Research Grant (Project ID 53090).
\clearpage
\appendix
\section{$\text{Model}f^{(II)}$ and the optical depth at reionization}
\begin{figure}[htbp]
    \centering
    \includegraphics[width=\textwidth]{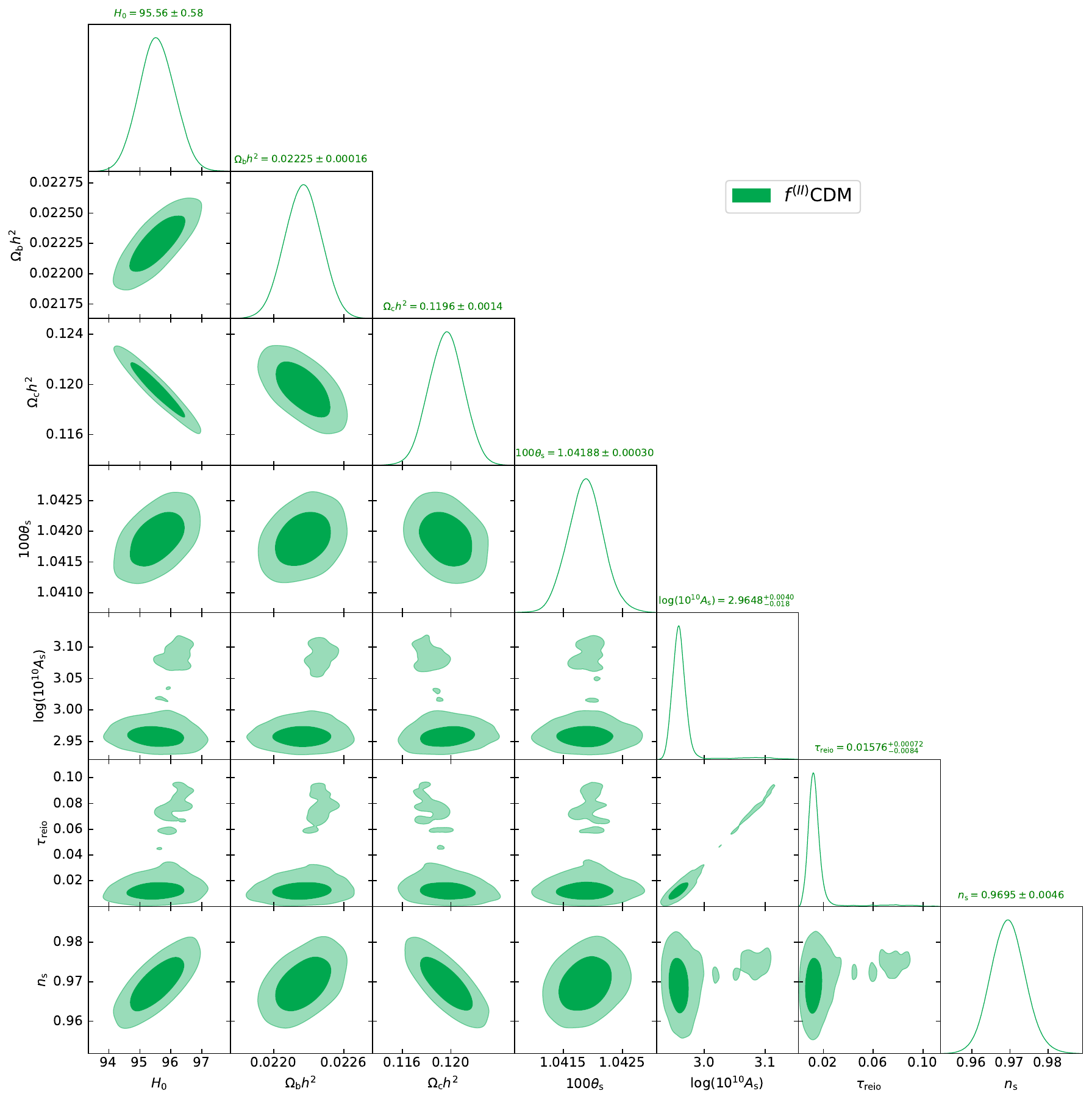}
    \caption{The 68\% (dark color) and 95\% (faint color) confidence level (CL) contours of the cosmological parameters for the $f^{(II)}$CDM model using high $\ell$+ low-$\ell$ TT data.}
    \label{fig:Planck+lowT}
\end{figure}

The inclusion of the low-$\ell$ TT data pushes the $\tau_{re}$ value to lower values $\sim 0.016$ not compatible with physical values, see Fig. \ref{fig:Planck+lowT}. Similarly for $A_s$ value due to the degeneracy between $\tau_{re}$ and $A_s$. 
\clearpage

\begin{figure}[htbp]
    \centering
    \includegraphics[width=\textwidth]{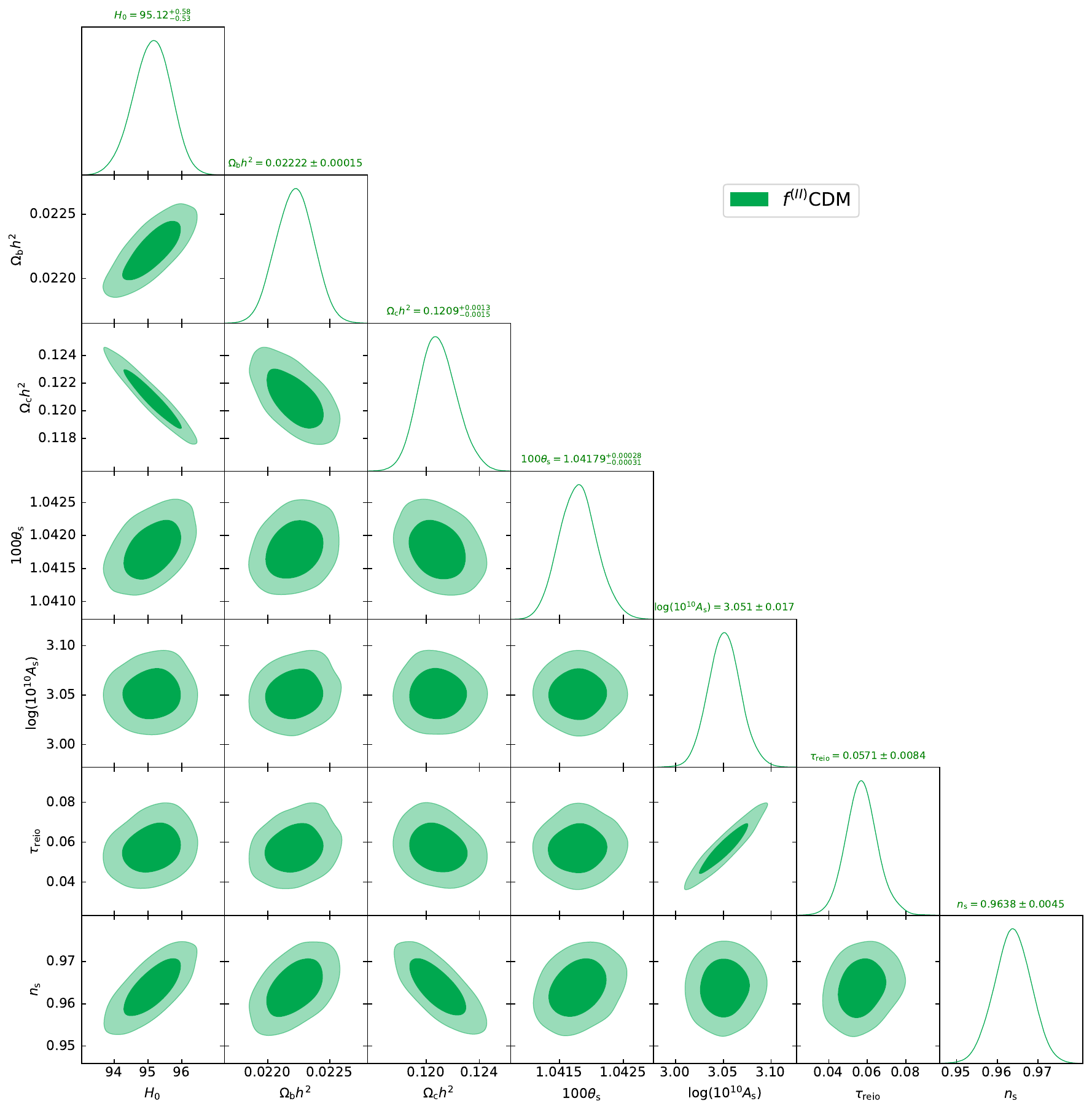}
    \caption{The 68\% (dark color) and 95\% (faint color) confidence level (CL) contours of the cosmological parameters for the $f^{(II)}$CDM model using high $\ell$+ low-$\ell$ EE data.}
    \label{fig:Planck+lowE}
\end{figure}

The inclusion of the low-$\ell$ EE data pushes the $\tau_{re}$ value to values $\sim 0.0564$ compatible with physical values, see Fig. \ref{fig:Planck+lowE}.
\clearpage

\begin{figure}[htbp]
    \centering
    \includegraphics[width=\textwidth]{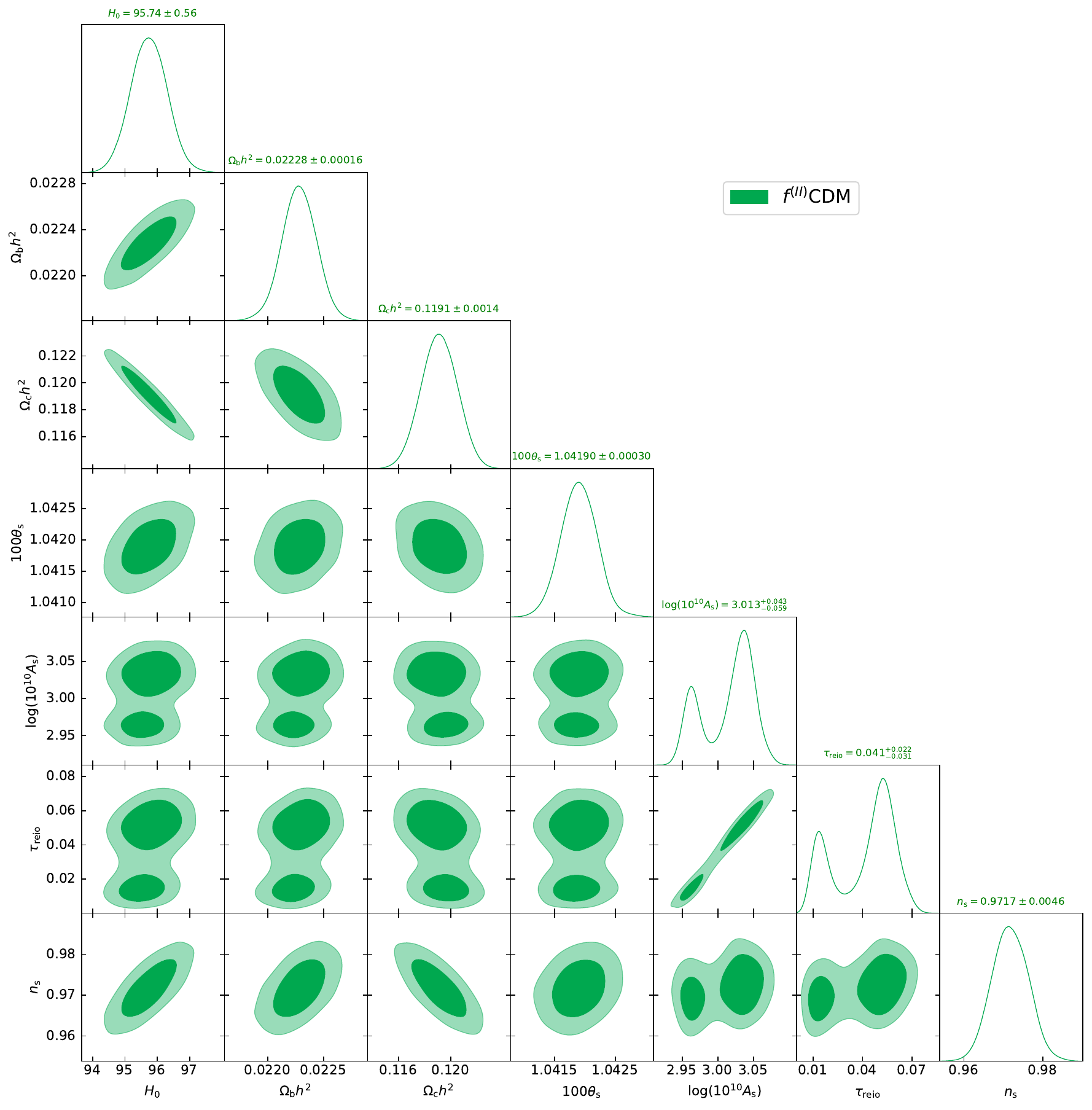}
    \caption{The 68\% (dark color) and 95\% (faint color) confidence level (CL) contours of the cosmological parameters for the $f^{(II)}$CDM model using high $\ell$ + low-$\ell$ TT + low-$\ell$ EE.}
    \label{fig:Planck+lowT+lowE}
\end{figure}

The inclusion of the low-$\ell$ TT+EE data pushes the $\tau_{re}$ value to double peak values $\sim 0.01$ and $\sim 0.041$ which shows the tension between low-$\ell$ TT  and low-$\ell$ EE datasets, see Fig. \ref{fig:Planck+lowT+lowE}.

\begin{figure}[htbp]
    \centering
    \includegraphics[width=\textwidth]{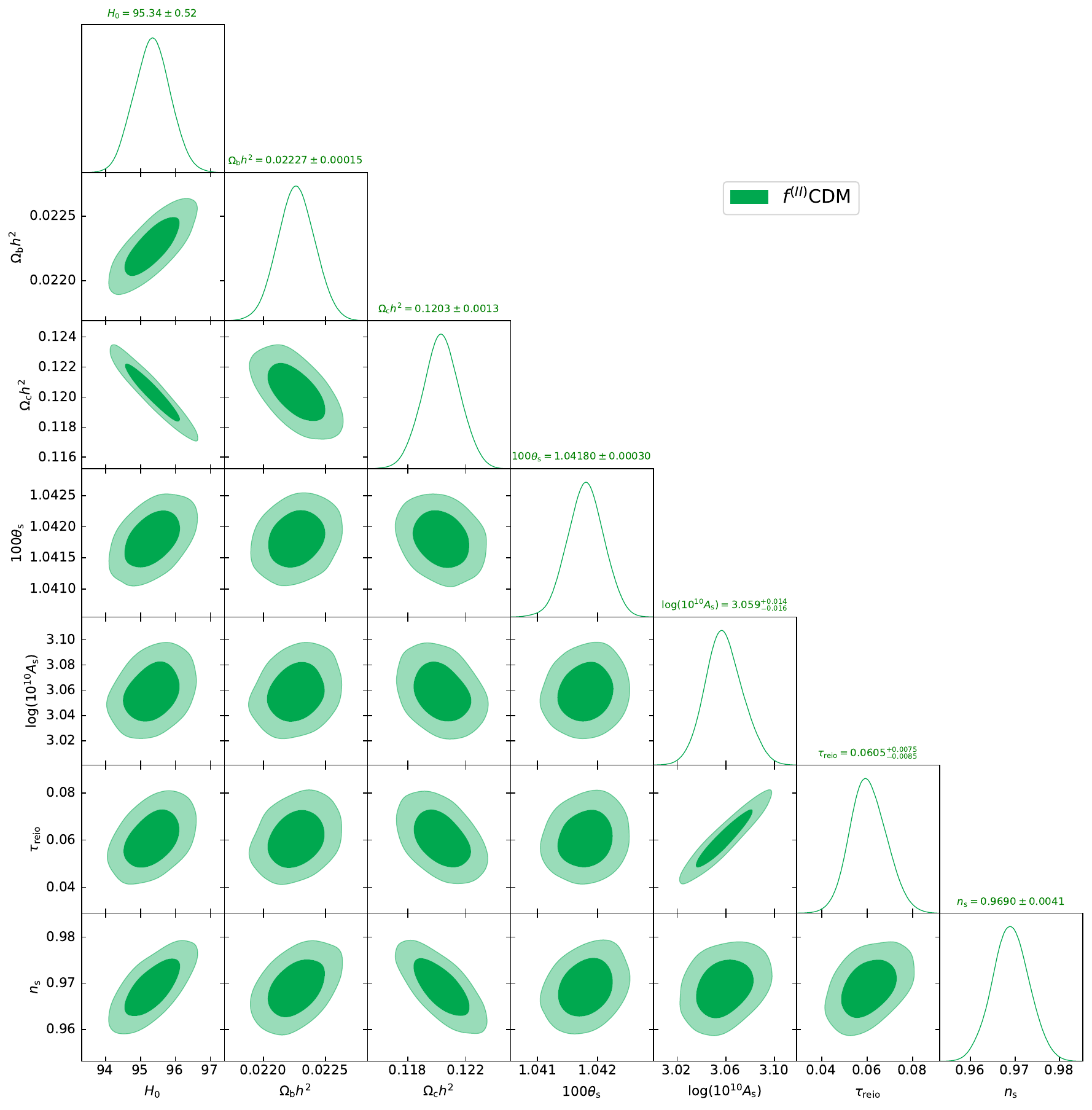}
    \caption{The 68\% (dark color) and 95\% (faint color) confidence level (CL) contours of the cosmological parameters for the $f^{(II)}$CDM model using high $\ell$ + low-$\ell$ TT + low-$\ell$ EE + lensing data.}
    \label{fig:Planck+lensing}
\end{figure}

\begin{figure}[htbp]
    \centering
    \includegraphics[width=\textwidth]{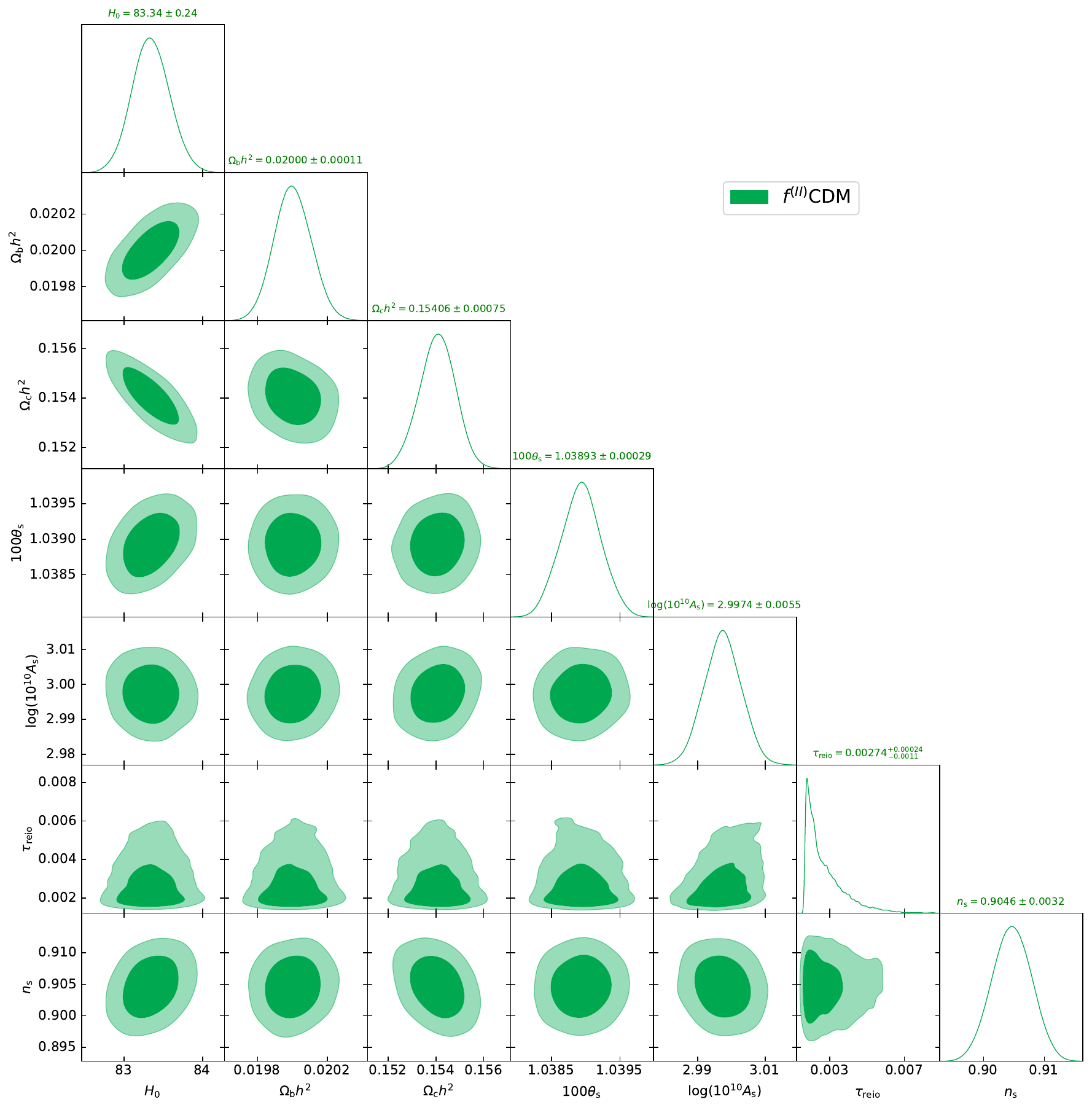}
    \caption{The 68\% (dark color) and 95\% (faint color) confidence level (CL) contours of the cosmological parameters for the $f^{(II)}$CDM model using high $\ell$ + low-$\ell$ TT + low-$\ell$ EE + lensing data+ DESI DR2 BAO.}
    \label{fig:Planck+lensing+DESI}
\end{figure}
\begin{figure}[htbp]
    \centering
    \includegraphics[width=\textwidth]{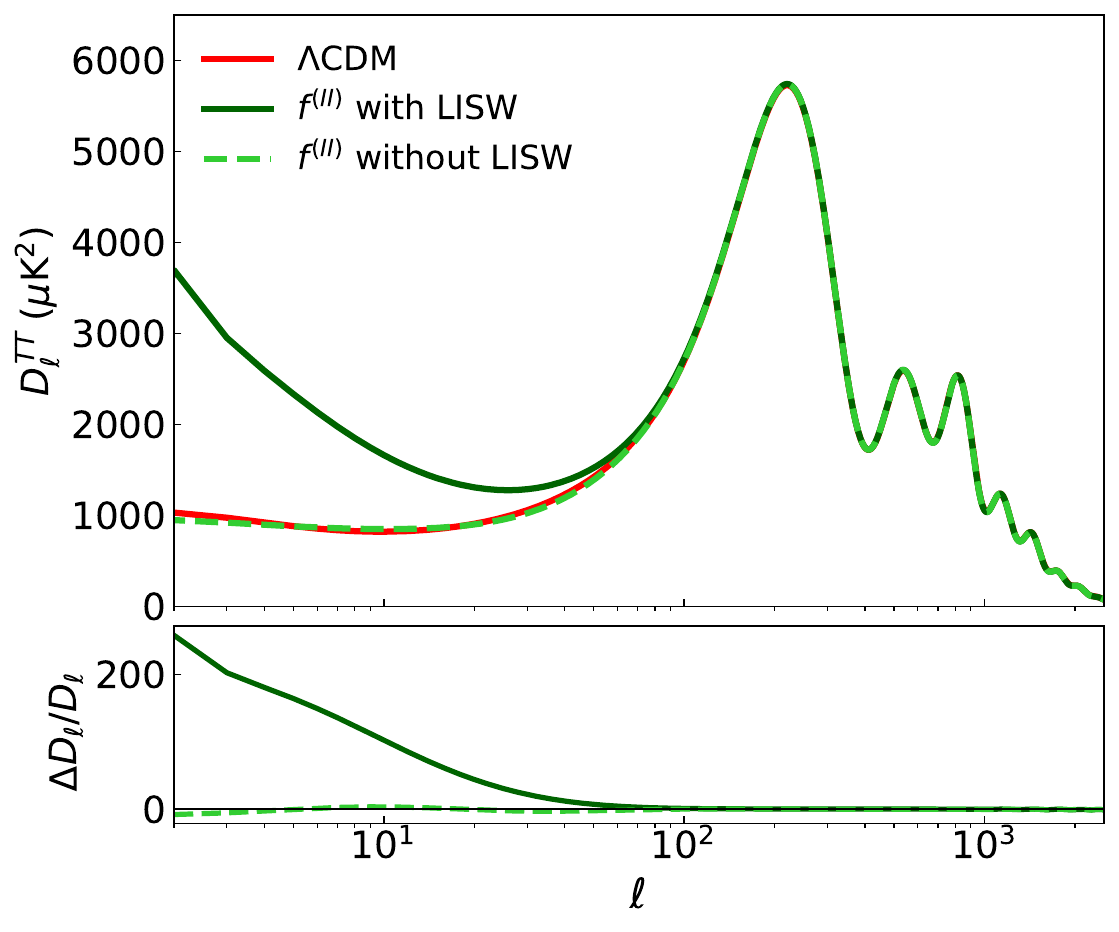}
    \caption{ $TT$ angular power spectra with and without the late Integrated Sachs–Wolfe (LISW) contribution for $f^{(II)}$ model. The enhanced LISW signal significantly modifies the low-$\ell$ $TT$ anisotropies, providing the physical mechanism by which the low-$\ell$ \textit{Planck} $TT$ likelihood drives the preferred values of $A_s$ and $\tau_{\rm reion}$, despite the large cosmic variance on these scales.
 }
    \label{fig:LISW}
\end{figure}
\clearpage

\bibliographystyle{apsrev4-1}
\bibliography{main}

@article{Planck:2018vyg,
    author = "Aghanim, N. and others",
    collaboration = "Planck",
    title = "{Planck 2018 results. VI. Cosmological parameters}",
    eprint = "1807.06209",
    archivePrefix = "arXiv",
    primaryClass = "astro-ph.CO",
    doi = "10.1051/0004-6361/201833910",
    journal = "Astron. Astrophys.",
    volume = "641",
    pages = "A6",
    year = "2020",
    note = "[Erratum: Astron.Astrophys. 652, C4 (2021)]"
}

@article{Menci:2024hop,
    author = "Menci, Nicola and Sen, Anjan Ananda and Castellano, Marco",
    title = "{The Excess of JWST Bright Galaxies: A Possible Origin in the Ground State of Dynamical Dark Energy in the Light of DESI 2024 Data}",
    eprint = "2410.22940",
    archivePrefix = "arXiv",
    primaryClass = "astro-ph.CO",
    doi = "10.3847/1538-4357/ad8d5b",
    journal = "Astrophys. J.",
    volume = "976",
    number = "2",
    pages = "227",
    year = "2024"
}

@book{Aldrovandi:2013wha,
    author = "Aldrovandi, Ruben and Pereira, Jos{\'e} Geraldo",
    title = "{Teleparallel Gravity}: {An Introduction}",
    doi = "10.1007/978-94-007-5143-9",
    isbn = "978-94-007-5142-2, 978-94-007-5143-9",
    publisher = "Springer",
    year = "2013"
}

@article{Cai:2026swf,
    author = "Cai, Rong-Gen and Wang, Shao-Jiang",
    title = "{The Hubble Tension: A Decade Review}",
    eprint = "2606.20434",
    archivePrefix = "arXiv",
    primaryClass = "astro-ph.CO",
    doi = "10.1088/1674-4527/ae842f",
    journal = "Res. Astron. Astrophys.",
    volume = "26",
    number = "8",
    pages = "084011",
    year = "2026"
}

@article{Gomez-Valent:2025mfl,
    author = "G{\'o}mez-Valent, Adri{\`a} and Gonz{\'a}lez-Fuentes, Alex",
    title = "{Effective phantom divide crossing with standard and negative quintessence}",
    eprint = "2508.00621",
    archivePrefix = "arXiv",
    primaryClass = "astro-ph.CO",
    doi = "10.1016/j.physletb.2025.140096",
    journal = "Phys. Lett. B",
    volume = "872",
    pages = "140096",
    year = "2026"
}

@article{Nojiri:2017ncd,
    author = "Nojiri, S. and Odintsov, S. D. and Oikonomou, V. K.",
    title = "{Modified Gravity Theories on a Nutshell: Inflation, Bounce and Late-time Evolution}",
    eprint = "1705.11098",
    archivePrefix = "arXiv",
    primaryClass = "gr-qc",
    reportNumber = "PHYS.REPT.-692-(2017)-1-104, Phys.Rept. 692 (2017) 1-104",
    doi = "10.1016/j.physrep.2017.06.001",
    journal = "Phys. Rept.",
    volume = "692",
    pages = "1--104",
    year = "2017"
}

@article{Li:2010cg,
    author = "Li, Baojiu and Sotiriou, Thomas P. and Barrow, John D.",
    title = "{$f(T)$ gravity and local Lorentz invariance}",
    eprint = "1010.1041",
    archivePrefix = "arXiv",
    primaryClass = "gr-qc",
    doi = "10.1103/PhysRevD.83.064035",
    journal = "Phys. Rev. D",
    volume = "83",
    pages = "064035",
    year = "2011"
}

@book{CANTATA:2021asi,
    author = "Akrami, Yashar and others",
    editor = "Saridakis, Emmanuel N. and Lazkoz, Ruth and Salzano, Vincenzo and Vargas Moniz, Paulo and Capozziello, Salvatore and Beltr{\'a}n Jim{\'e}nez, Jose and De Laurentis, Mariafelicia and Olmo, Gonzalo J.",
    collaboration = "CANTATA",
    title = "{Modified Gravity and Cosmology. An Update by the CANTATA Network}",
    eprint = "2105.12582",
    archivePrefix = "arXiv",
    primaryClass = "gr-qc",
    doi = "10.1007/978-3-030-83715-0",
    isbn = "978-3-030-83714-3, 978-3-030-83717-4, 978-3-030-83715-0",
    publisher = "Springer",
    year = "2021"
}

@article{Riess:2016jrr,
    author = "Riess, Adam G. and others",
    title = "{A 2.4{\%} Determination of the Local Value of the Hubble Constant}",
    eprint = "1604.01424",
    archivePrefix = "arXiv",
    primaryClass = "astro-ph.CO",
    doi = "10.3847/0004-637X/826/1/56",
    journal = "Astrophys. J.",
    volume = "826",
    number = "1",
    pages = "56",
    year = "2016"
}

@article{DES:2024jxu,
    author = "Abbott, T. M. C. and others",
    collaboration = "DES",
    title = "{The Dark Energy Survey: Cosmology Results with {\ensuremath{\sim}}1500 New High-redshift Type Ia Supernovae Using the Full 5 yr Data Set}",
    eprint = "2401.02929",
    archivePrefix = "arXiv",
    primaryClass = "astro-ph.CO",
    reportNumber = "FERMILAB-PUB-23-0821-PPD, DES-2023-805",
    doi = "10.3847/2041-8213/ad6f9f",
    journal = "Astrophys. J. Lett.",
    volume = "973",
    number = "1",
    pages = "L14",
    year = "2024"
}

@article{Brout:2022vxf,
    author = "Brout, Dillon and others",
    title = "{The Pantheon+ Analysis: Cosmological Constraints}",
    eprint = "2202.04077",
    archivePrefix = "arXiv",
    primaryClass = "astro-ph.CO",
    doi = "10.3847/1538-4357/ac8e04",
    journal = "Astrophys. J.",
    volume = "938",
    number = "2",
    pages = "110",
    year = "2022"
}

@article{Rubin:2023jdq,
    author = "Rubin, David and others",
    title = "{Union Through UNITY: Cosmology with 2,000 SNe Using a Unified Bayesian Framework}",
    eprint = "2311.12098",
    archivePrefix = "arXiv",
    primaryClass = "astro-ph.CO",
    doi = "10.3847/1538-4357/adc0a5",
    journal = "Astrophys. J.",
    volume = "986",
    number = "2",
    pages = "231",
    year = "2025"
}

@article{Awad:2017yod,
    author = "Awad, A. and El Hanafy, W. and Nashed, G. G. L. and Saridakis, Emmanuel N.",
    title = "{Phase Portraits of general f(T) Cosmology}",
    eprint = "1710.10194",
    archivePrefix = "arXiv",
    primaryClass = "gr-qc",
    doi = "10.1088/1475-7516/2018/02/052",
    journal = "JCAP",
    volume = "02",
    pages = "052",
    year = "2018"
}

@article{Hashim:2021pkq,
    author = "Hashim, Mahmoud and El-Zant, Amr A. and El Hanafy, Waleed and Golovnev, Alexey",
    title = "{Toward a concordance teleparallel cosmology. Part~II. Linear perturbation}",
    eprint = "2104.08311",
    archivePrefix = "arXiv",
    primaryClass = "astro-ph.CO",
    doi = "10.1088/1475-7516/2021/07/053",
    journal = "JCAP",
    volume = "07",
    pages = "053",
    year = "2021"
}

@article{DESI:2025zgx,
    author = "Abdul Karim, M. and others",
    collaboration = "DESI",
    title = "{DESI DR2 results. II. Measurements of baryon acoustic oscillations and cosmological constraints}",
    eprint = "2503.14738",
    archivePrefix = "arXiv",
    primaryClass = "astro-ph.CO",
    reportNumber = "FERMILAB-PUB-25-0169-PPD",
    doi = "10.1103/tr6y-kpc6",
    journal = "Phys. Rev. D",
    volume = "112",
    number = "8",
    pages = "083515",
    year = "2025"
}

@article{Golovnev:2018wbh,
    author = "Golovnev, Alexey and Koivisto, Tomi",
    title = "{Cosmological perturbations in modified teleparallel gravity models}",
    eprint = "1808.05565",
    archivePrefix = "arXiv",
    primaryClass = "gr-qc",
    reportNumber = "NORDITA 2018-073",
    doi = "10.1088/1475-7516/2018/11/012",
    journal = "JCAP",
    volume = "11",
    pages = "012",
    year = "2018"
}

@article{Planck:2018lbu,
    author = "Aghanim, N. and others",
    collaboration = "Planck",
    title = "{Planck 2018 results. VIII. Gravitational lensing}",
    eprint = "1807.06210",
    archivePrefix = "arXiv",
    primaryClass = "astro-ph.CO",
    doi = "10.1051/0004-6361/201833886",
    journal = "Astron. Astrophys.",
    volume = "641",
    pages = "A8",
    year = "2020"
}

@ARTICLE{2011JCAP...07..034B,
       author = {{Blas}, Diego and {Lesgourgues}, Julien and {Tram}, Thomas},
        title = "{The Cosmic Linear Anisotropy Solving System (CLASS). Part II: Approximation schemes}",
      journal = {Journal of Cosmology and Astroparticle Physics},
         year = 2011,
        month = jul,
       volume = {2011},
       number = {7},
          eid = {034},
        pages = {034},
          doi = {10.1088/1475-7516/2011/07/034},
archivePrefix = {arXiv},
       eprint = {1104.2933},
 primaryClass = {astro-ph.CO},
       adsurl = {https://ui.adsabs.harvard.edu/abs/2011JCAP...07..034B}
}

@article{Lewis:2019xzd,
    author = "Lewis, Antony",
    title = "{GetDist: a Python package for analysing Monte Carlo samples}",
    eprint = "1910.13970",
    archivePrefix = "arXiv",
    primaryClass = "astro-ph.IM",
    doi = "10.1088/1475-7516/2025/08/025",
    journal = "JCAP",
    volume = "08",
    pages = "025",
    year = "2025"
}

@article{Torrado:2020dgo,
    author = "Torrado, Jesus and Lewis, Antony",
    title = "{Cobaya: Code for Bayesian Analysis of hierarchical physical models}",
    eprint = "2005.05290",
    archivePrefix = "arXiv",
    primaryClass = "astro-ph.IM",
    reportNumber = "TTK-20-15",
    doi = "10.1088/1475-7516/2021/05/057",
    journal = "JCAP",
    volume = "05",
    pages = "057",
    year = "2021"
}

@article{Ferreira:2025lrd,
    author = "Ferreira, Elisa G. M. and McDonough, Evan and Balkenhol, Lennart and Kallosh, Renata and Knox, Lloyd and Linde, Andrei",
    title = "{BAO-CMB tension and implications for inflation}",
    eprint = "2507.12459",
    archivePrefix = "arXiv",
    primaryClass = "astro-ph.CO",
    doi = "10.1103/lq71-b84v",
    journal = "Phys. Rev. D",
    volume = "113",
    number = "4",
    pages = "043524",
    year = "2026"
}

@article{Maluf:2013gaa,
    author = "Maluf, J. W.",
    title = "{The teleparallel equivalent of general relativity}",
    eprint = "1303.3897",
    archivePrefix = "arXiv",
    primaryClass = "gr-qc",
    doi = "10.1002/andp.201200272",
    journal = "Annalen Phys.",
    volume = "525",
    pages = "339--357",
    year = "2013"
}

@article{Li:2011wu,
    author = "Li, Baojiu and Sotiriou, Thomas P. and Barrow, John D.",
    title = "{Large-scale Structure in f(T) Gravity}",
    eprint = "1103.2786",
    archivePrefix = "arXiv",
    primaryClass = "astro-ph.CO",
    doi = "10.1103/PhysRevD.83.104017",
    journal = "Phys. Rev. D",
    volume = "83",
    pages = "104017",
    year = "2011"
}

@article{Hashim:2020sez,
    author = "Hashim, Mahmoud and El Hanafy, Waleed and Golovnev, Alexey and El-Zant, Amr A.",
    title = "{Toward a concordance teleparallel cosmology. Part I. Background dynamics}",
    eprint = "2010.14964",
    archivePrefix = "arXiv",
    primaryClass = "astro-ph.CO",
    doi = "10.1088/1475-7516/2021/07/052",
    journal = "JCAP",
    volume = "07",
    pages = "052",
    year = "2021"
}

@inproceedings{DeAndrade:2000sf,
    author = "De Andrade, V. C. and Guillen, L. C. T. and Pereira, J. G.",
    title = "{Teleparallel gravity: An Overview}",
    booktitle = "{9th Marcel Grossmann Meeting on Recent Developments in Theoretical and Experimental General Relativity, Gravitation and Relativistic Field Theories (MG 9)}",
    eprint = "gr-qc/0011087",
    archivePrefix = "arXiv",
    month = "11",
    year = "2000"
}

@BOOK{2013tgif.book.....A,
       author = {{Aldrovandi}, Ruben and {Pereira}, Jos{\'e} Geraldo},
        title = "{Teleparallel Gravity: An Introduction}",
         year = 2013,
       volume = {173},
          doi = {10.1007/978-94-007-5143-9},
       adsurl = {https://ui.adsabs.harvard.edu/abs/2013tgif.book.....A}
}

@article{Bahamonde:2021gfp,
    author = "Bahamonde, Sebastian and Dialektopoulos, Konstantinos F. and Escamilla-Rivera, Celia and Farrugia, Gabriel and Gakis, Viktor and Hendry, Martin and Hohmann, Manuel and Levi Said, Jackson and Mifsud, Jurgen and Di Valentino, Eleonora",
    title = "{Teleparallel gravity: from theory to cosmology}",
    eprint = "2106.13793",
    archivePrefix = "arXiv",
    primaryClass = "gr-qc",
    doi = "10.1088/1361-6633/ac9cef",
    journal = "Rept. Prog. Phys.",
    volume = "86",
    number = "2",
    pages = "026901",
    year = "2023"
}

@article{Nesseris:2013jea,
    author = "Nesseris, S. and Basilakos, S. and Saridakis, E. N. and Perivolaropoulos, L.",
    title = "{Viable $f(T)$ models are practically indistinguishable from $\Lambda$CDM}",
    eprint = "1308.6142",
    archivePrefix = "arXiv",
    primaryClass = "astro-ph.CO",
    reportNumber = "IFT-UAM-CSIC-13-091",
    doi = "10.1103/PhysRevD.88.103010",
    journal = "Phys. Rev. D",
    volume = "88",
    pages = "103010",
    year = "2013"
}

@article{Efstathiou:1998xx,
    author = "Efstathiou, G. and Bond, J. R.",
    title = "{Cosmic confusion: Degeneracies among cosmological parameters derived from measurements of microwave background anisotropies}",
    eprint = "astro-ph/9807103",
    archivePrefix = "arXiv",
    reportNumber = "GPE-98-7",
    doi = "10.1046/j.1365-8711.1999.02274.x",
    journal = "Mon. Not. Roy. Astron. Soc.",
    volume = "304",
    pages = "75--97",
    year = "1999"
}

@ARTICLE{Ludwick:2017tox,
  author       = {Ludwick, Kevin J.},
  title        = {{The viability of phantom dark energy: A review}},
  journal      = {Mod. Phys. Lett.},
  volume       = A32,
  year         = 2017,
  number       = 28,
  pages        = 1730025,
  doi          = {10.1142/S0217732317300257},
  eprint       = {1708.06981},
  archiveprefix= {arXiv},
  primaryclass = {astro-ph.CO},
}

@article{Santos:2022atq,
    author = "Santos, F. B. M. dos",
    title = "{Updating constraints on phantom crossing f(T) gravity}",
    eprint = "2211.16370",
    archivePrefix = "arXiv",
    primaryClass = "astro-ph.CO",
    doi = "10.1088/1475-7516/2023/06/039",
    journal = "JCAP",
    volume = "06",
    pages = "039",
    year = "2023"
}

@article{Gomez-Valent:2024ejh,
    author = "G{\'o}mez-Valent, Adria and Sol{\`a} Peracaula, Joan",
    title = "{Composite dark energy and the cosmological tensions}",
    eprint = "2412.15124",
    archivePrefix = "arXiv",
    primaryClass = "astro-ph.CO",
    doi = "10.1016/j.physletb.2025.139391",
    journal = "Phys. Lett. B",
    volume = "864",
    pages = "139391",
    year = "2025"
}

@article{Soriano:2025gxd,
    author = "Soriano, Jorge F. and Wohlberg, Shimon and Anchordoqui, Luis A.",
    title = "{New insights on a sign-switching {\ensuremath{\Lambda}}}",
    eprint = "2502.19239",
    archivePrefix = "arXiv",
    primaryClass = "astro-ph.CO",
    doi = "10.1016/j.dark.2025.101911",
    journal = "Phys. Dark Univ.",
    volume = "48",
    pages = "101911",
    year = "2025"
}

@article{BOSS:2014hhw,
    author = "Aubourg, {\'E}ric and others",
    collaboration = "BOSS",
    title = "{Cosmological implications of baryon acoustic oscillation measurements}",
    eprint = "1411.1074",
    archivePrefix = "arXiv",
    primaryClass = "astro-ph.CO",
    doi = "10.1103/PhysRevD.92.123516",
    journal = "Phys. Rev. D",
    volume = "92",
    number = "12",
    pages = "123516",
    year = "2015"
}

@article{BOSS:2016wmc,
    author = "Alam, Shadab and others",
    collaboration = "BOSS",
    title = "{The clustering of galaxies in the completed SDSS-III Baryon Oscillation Spectroscopic Survey: cosmological analysis of the DR12 galaxy sample}",
    eprint = "1607.03155",
    archivePrefix = "arXiv",
    primaryClass = "astro-ph.CO",
    doi = "10.1093/mnras/stx721",
    journal = "Mon. Not. Roy. Astron. Soc.",
    volume = "470",
    number = "3",
    pages = "2617--2652",
    year = "2017"
}

@article{eBOSS:2019ytm,
    author = "de Sainte Agathe, Victoria and others",
    collaboration = "eBOSS",
    title = "{Baryon acoustic oscillations at z = 2.34 from the correlations of Ly$\alpha$ absorption in eBOSS DR14}",
    eprint = "1904.03400",
    archivePrefix = "arXiv",
    primaryClass = "astro-ph.CO",
    doi = "10.1051/0004-6361/201935638",
    journal = "Astron. Astrophys.",
    volume = "629",
    pages = "A85",
    year = "2019"
}

@article{eBOSS:2020xwt,
    author = "Wang, Yuting and others",
    collaboration = "eBOSS",
    title = "{The clustering of the SDSS-IV extended Baryon Oscillation Spectroscopic Survey DR16 luminous red galaxy and emission line galaxy samples: cosmic distance and structure growth measurements using multiple tracers in configuration space}",
    eprint = "2007.09010",
    archivePrefix = "arXiv",
    primaryClass = "astro-ph.CO",
    doi = "10.1093/mnras/staa2593",
    journal = "Mon. Not. Roy. Astron. Soc.",
    volume = "498",
    number = "3",
    pages = "3470--3483",
    year = "2020"
}

@article{DESI:2024aqx,
    author = "Calderon, R. and others",
    collaboration = "DESI",
    title = "{DESI 2024: reconstructing dark energy using crossing statistics with DESI DR1 BAO data}",
    eprint = "2405.04216",
    archivePrefix = "arXiv",
    primaryClass = "astro-ph.CO",
    doi = "10.1088/1475-7516/2024/10/048",
    journal = "JCAP",
    volume = "10",
    pages = "048",
    year = "2024"
}

@article{Menci:2024rbq,
    author = "Menci, Nicola and Adil, Shahnawaz A. and Mukhopadhyay, Upala and Sen, Anjan A. and Vagnozzi, Sunny",
    title = "{Negative cosmological constant in the dark energy sector: tests from JWST photometric and spectroscopic observations of high-redshift galaxies}",
    eprint = "2401.12659",
    archivePrefix = "arXiv",
    primaryClass = "astro-ph.CO",
    doi = "10.1088/1475-7516/2024/07/072",
    journal = "JCAP",
    volume = "07",
    pages = "072",
    year = "2024"
}

@UNPUBLISHED{Hashim:2026M3,
  AUTHOR =       "Hashim, Mahmoud and Di Valentino, Eleonora and Levi Said, Jackson and El Hanafy, Waleed",
  TITLE =        "{New Exponential Infrared $f(T)$ Gravity}",
  NOTE =         {In progress},
  year =         {2026},
  month =        {September},
}

@UNPUBLISHED{Bedair:2026M4,
  AUTHOR =       "Bedair, Mohamed and Hashim, Mahmoud and Di Valentino, Eleonora and Levi Said, Jackson and El Hanafy, Waleed",
  TITLE =        "{A Six-Parameter Teleparallel Cosmology Beyond $\Lambda$CDM: Sign-Changing Torsional Dark Energy and Implications for $H_0$}",
  NOTE =         {In progress},
  year =         {2026},
  month =        {September},
}

@article{Escamilla-Rivera:2024sae,
    author = "Escamilla-Rivera, Celia and Sandoval-Orozco, Rodrigo",
    title = "{f(T) gravity after DESI Baryon acoustic oscillation and DES supernovae 2024 data}",
    eprint = "2405.00608",
    archivePrefix = "arXiv",
    primaryClass = "astro-ph.CO",
    doi = "10.1016/j.jheap.2024.05.005",
    journal = "JHEAp",
    volume = "42",
    pages = "217--221",
    year = "2024"
}

@article{SupernovaSearchTeam:1998fmf,
    author = "Riess, Adam G. and others",
    collaboration = "Supernova Search Team",
    title = "{Observational evidence from supernovae for an accelerating universe and a cosmological constant}",
    eprint = "astro-ph/9805201",
    archivePrefix = "arXiv",
    doi = "10.1086/300499",
    journal = "Astron. J.",
    volume = "116",
    pages = "1009--1038",
    year = "1998"
}

@article{SupernovaCosmologyProject:1998vns,
    author = "Perlmutter, S. and others",
    collaboration = "Supernova Cosmology Project",
    title = "{Measurements of $\Omega$ and $\Lambda$ from 42 High Redshift Supernovae}",
    eprint = "astro-ph/9812133",
    archivePrefix = "arXiv",
    reportNumber = "LBNL-41801, LBL-41801",
    doi = "10.1086/307221",
    journal = "Astrophys. J.",
    volume = "517",
    pages = "565--586",
    year = "1999"
}

@article{Heisenberg:2023lru,
    author = "Heisenberg, Lavinia",
    title = "{Review on f(Q) gravity}",
    eprint = "2309.15958",
    archivePrefix = "arXiv",
    primaryClass = "gr-qc",
    doi = "10.1016/j.physrep.2024.02.001",
    journal = "Phys. Rept.",
    volume = "1066",
    pages = "1--78",
    year = "2024"
}

@UNPUBLISHED{Hashim:2026M2,
  AUTHOR =       "Hashim, Mahmoud and Di Valentino, Eleonora and Levi Said, Jackson and El Hanafy, Waleed",
  TITLE =       "{Extended Exponential Infrared $f(T)$ Gravity}",
  NOTE =         {In progress},
  year =         {2026},
  month =        {September},
}

@article{Akarsu:2021fol,
    author = {Akarsu, {\"O}zg{\"u}r and Kumar, Suresh and {\"O}z{\"u}lker, Emre and Vazquez, J. Alberto},
    title = "{Relaxing cosmological tensions with a sign switching cosmological constant}",
    eprint = "2108.09239",
    archivePrefix = "arXiv",
    primaryClass = "astro-ph.CO",
    doi = "10.1103/PhysRevD.104.123512",
    journal = "Phys. Rev. D",
    volume = "104",
    number = "12",
    pages = "123512",
    year = "2021"
}

@article{Akarsu:2024nas,
    author = "Akarsu, Ozgur and Bulduk, Bilal and De Felice, Antonio and Kat{\i}rc{\i}, Nihan and Uzun, N. Merve",
    title = "{Unexplored regions in teleparallel f(T) gravity: Sign-changing dark energy density}",
    eprint = "2410.23068",
    archivePrefix = "arXiv",
    primaryClass = "gr-qc",
    reportNumber = "YITP-24-119",
    doi = "10.1103/1xd4-k91h",
    journal = "Phys. Rev. D",
    volume = "112",
    number = "8",
    pages = "083532",
    year = "2025"
}

@article{Hashim:2026yoy,
    author = "Hashim, Mahmoud and Di Valentino, Eleonora and Levi Said, Jackson and El Hanafy, Waleed",
    title = "{Cosmological Viability of Exponential Infrared $f(T)$ Gravity}",
    eprint = "2606.31324",
    archivePrefix = "arXiv",
    primaryClass = "astro-ph.CO",
    month = "6",
    year = "2026"
}
\end{document}